\documentclass[twocolumn,superscriptaddress,floatfix,noeprint]{revtex4-2}

\usepackage{graphicx,makecell}
\usepackage{amsmath,amsfonts,amssymb}
\usepackage{mathtools}
\usepackage{mathrsfs}
\usepackage{bm,dsfont}
\usepackage[dvipsnames]{xcolor}
\usepackage[colorlinks=true, citecolor=Green, linkcolor=BrickRed, urlcolor=NavyBlue]{hyperref}
\usepackage[all]{hypcap}

\graphicspath{{figures/}}

\DeclareMathOperator{\sgn}{sgn}

\begin{document}

\title{Floquet-Bloch Theory for Nonperturbative Response to a Static Drive}

\author{Christophe De Beule}
\author{Steven Gassner}
\author{Spenser Talkington}
\author{E. J. Mele}
\affiliation{Department of Physics and Astronomy, University of Pennsylvania, Philadelphia, Pennsylvania 19104, USA}

\date{\today}

\begin{abstract}
We develop the Floquet-Bloch theory of noninteracting fermions on a periodic lattice
in the presence of a \emph{constant} electric field. As long as the field lies along a reciprocal lattice vector, time periodicity of the Bloch Hamiltonian is inherited from the evolution of momentum in the Brillouin zone. The corresponding Floquet quasienergies yield the Wannier-Stark ladder with interband couplings included to all orders. These results are compared to perturbative results where the lowest-order interband correction gives the field-induced polarization shift in terms of the electric susceptibility. Additionally, we investigate electronic transport by coupling the system to a bath within the Floquet-Keldysh formalism. We then study the breakdown of the band-projected theory from the onset of interband contributions and Zener resonances in the band-resolved currents. In particular, we consider the transverse quantum-geometric response in two spatial dimensions due to the Berry curvature. In the strong-field regime, the semiclassical theory predicts a plateau of the geometric response as a function of field strength. Here, we scrutinize the conditions under which the semiclassical results continue to hold in the quantum theory.
\end{abstract}

\maketitle

\section{Introduction}

A peculiar property of electrons in a crystal is that applying a constant direct current (dc) electric field gives rise to periodic motion known as Bloch oscillations. As the name suggests, this result was first derived by Felix Bloch in 1929 \cite{bloch_uber_1929}, although the oscillatory electron trajectories were only mentioned explicitly by Clarence Zener five years later \cite{zener_theory_1934}. Seemingly counterintuitive, this phenomenon follows naturally from the fact that in the presence of a periodic potential, momentum space is compactified to a torus, i.e., the first Brillouin zone (BZ). Since in a dc electric field $\bm{\mathcal E}$, momentum increases linearly with time, $\bm k(t) = \bm k(0) - e \bm{\mathcal E}t / \hbar$, this gives rise to oscillatory motion with characteristic Bloch frequency $\omega_B = e |\bm{\mathcal E}| a / \hbar$ where $a$ is the lattice constant \cite{esaki_superlattice_1970,kurz_bloch_1996,tsu_superlattice_2005}.

However, Bloch oscillations are typically inconsequential for electronic transport because most solid-state systems obey $\omega_B \tau \ll 1$ where $\tau$ is a typical momentum-relaxation time. In this case, an electron relaxes its momentum to equilibrium before a complete oscillation occurs. One approach to reach the regime $\omega_B \tau \gtrsim 1$ at reasonable field strengths is by means of artificial superlattices, such as quasi-one-dimensional (1D) semiconductor superlattices \cite{leo_observation_1992,mendez_wannierstark_1993,bouchard_bloch_1995} or two-dimensional (2D) moir\'e systems \cite{fahimniya_synchronizing_2021,vakhtel_bloch_2022,phong_quantum_2023,de_beule_roses_2023}, among others. Here the large lattice constant $a \sim 10 \; \text{nm}$ compresses the BZ which leads to an increase in $\omega_B$ such that full Bloch orbits can be executed before a scattering event occurs.

Experimental observations of Bloch oscillations in solid-state systems have therefore been limited to 1D semiconductor superlattices where a static electric field is applied along the superlattice direction and Bloch oscillations with $\omega_B$ in the THz regime are excited optically \cite{feldmann_optical_1992,leo_observation_1992,waschke_coherent_1993,ignatov_thz-field_1995,dekorsy_bloch_1995,savvidis_resonant_2004}. Moreover, the presence of Bloch oscillations also manifests itself in the dc electronic response through a negative differential conductance  \cite{esaki_superlattice_1970,sibille_observation_1990} for $\omega_B \tau \gtrsim 1$. Such a decrease in the current with increasing electric field strength is due to the Wannier-Stark (WS) localization of electrons in a static electric field \cite{shockley_stark_1972,krieger_time_1986,emin_existence_1987,voisin_observation_1988,nenciu_dynamics_1991}. Additionally, it has been recognized recently that the quantum geometry of Bloch bands, specifically the Berry curvature, gives rise to so-called geometric oscillations transverse to the electric field \cite{phong_quantum_2023}. Moreover, these geometric oscillations lead to a characteristic peak in the transverse differential conductance \cite{phong_quantum_2023,de_beule_roses_2023}, as well as peaks in the optical Hall conductivity that encode the quantum geometry of the band \cite{de_beule_berry_2023}. Furthermore, Bloch oscillations of cold atoms have been observed in optical lattices \cite{ben_dahan_bloch_1996,kolovsky_bloch_2004}. The latter are extremely clean systems such that $\omega_B \tau$ can be large even though the Bloch period can be up to $10$ orders of magnitude larger than in solid-state superlattices.

As an illustrative example, consider a one-dimensional crystal with a partially occupied energy band $E_k$. In this case, we have $\bm{\mathcal E} = \mathcal E \bm{\hat x}$ and the semiclassical electron trajectories are obtained from
\begin{equation}
    \frac{dx}{dt} = \frac{1}{\hbar} \frac{\partial E_{k(t)}}{\partial k} = -\frac{1}{e \mathcal E} \frac{dE_{k(t)}}{dt},
\end{equation}
with $-e$ the electron charge. The solution reads $x(t) = - E[k(t)] / e \mathcal E$ up to an additive constant, where the amplitude of the oscillation is given by $W / e |\mathcal E|$ for electronic bandwidth $W$. Hence the size of the Bloch orbit is proportional to $1/|\mathcal E|$, which is the previously mentioned WS localization. Complete orbits thus require that the size of the orbit be smaller than the mean free path, while only partial orbits are completed otherwise.

In higher spatial dimensions, the resulting trajectory in momentum space is generally not closed for an arbitrary direction of the electric field and eventually traces out the entire BZ. In this case, generic trajectories in both momentum and real space are oscillatory but not periodic. Electron motion is only periodic when the electric field lies along a reciprocal lattice vector $\bm g$, or equivalently when it is perpendicular to a lattice plane. In this case, the frequency is given by $\Omega =  2\pi \omega_B a / |\bm{g}|$. The corresponding real-space semiclassical trajectories generally consist of three parts: an oscillatory motion with components along the field direction, a transverse oscillatory motion originating from the anisotropy of the energy band as well as the anomalous velocity, and a transverse drift \cite{phong_quantum_2023}. The latter originates from velocity components not accelerated by the field and only contributes to the net transport current when time-reversal symmetry is broken. It corresponds to a drift of the guiding center of the Bloch orbit and contains a topological term proportional to the Chern number of the band.

So far we have tacitly assumed the adiabatic limit where the energy band is sufficiently isolated from other bands. In this limit, we can neglect interband Zener transitions and only consider the dynamics in the occupied band. At lowest order, interband corrections induce a field-dependent polarization shift \cite{gao_field_2014} which corresponds to the electric susceptibility \cite{komissarov_quantum_2023}. However, this term only contributes to transport in systems where spatial inversion symmetry and time-reversal symmetry are simultaneously broken. In this work, we consider interband corrections for electric transport beyond perturbation theory in $|\bm{\mathcal E}|$. We accomplish this by treating the dynamics exactly with Floquet theory for a \emph{commensurate} constant electric field $\bm{\mathcal E} \parallel \bm g$ \cite{gluck_calculation_1998,gluck_resonant_2000,gluck_wannierstark_2002,lee_direct_2015,kim_anomalous_2020}. In this case, we know that the band-projected semiclassical dynamics is periodic since momentum space is defined on a torus whose generators are reciprocal lattice vectors. The quantum theory inherits this feature from minimal substitution $\bm k \rightarrow \bm k + e \bm A(t)$ in temporal gauge (also known as velocity gauge) for a uniform electric field. As such, we will show that the Bloch Hamiltonian becomes time periodic up to a unitary transformation for commensurate fields. Importantly, by treating a \emph{static} drive in temporal gauge, one finds that the longitudinal momentum becomes a gauge degree of freedom which can be absorbed in the time origin. Hence, the original \emph{time-independent} problem in $D$ spatial dimensions is mapped to a \emph{time-dependent} problem in $D-1$ spatial dimensions. This is similar to how a periodic drive in $D$ dimensions can be mapped to a time-independent problem on the Floquet ladder in $D+1$ dimensions \cite{gomez-leon_floquet-bloch_2013}. In fact, here we are trading one physical spatial dimension, defined by the direction of the electric field, with a synthetic Floquet dimension along which translation symmetry is broken by the electrostatic potential.

This paper is further structured as follows. In Section \ref{sec:floquet}, we introduce the Floquet formalism for the nonperturbative treatment of a static uniform electric field applied to a periodic lattice of free fermions. Special care is taken to account for the sublattices by a modified Floquet \emph{ansatz} yielding the Floquet Hamiltonian whose quasienergies give the WS ladder. Then in Section \ref{sec:transport}, we couple the clean system to a reservoir with the Floquet-Keldysh formalism and calculate the band-resolved charge currents. Next, in Section \ref{sec:examples} we apply the transport theory to examples in one and two spatial dimensions. Specifically, we study the onset of interband Zener resonances in a 1D dimer chain, as well as the current anisotropy and transverse geometric response for the honeycomb lattice in 2D with sublattice and Haldane mass terms. We finally present our conclusions in Section \ref{sec:conclusions}. Unless stated otherwise, we set $\hbar = 1$ from now on.

\section{Floquet theory of a crystal in a constant electric field} \label{sec:floquet}

In this section, we introduce the Floquet formalism for the nonperturbative treatment of a static uniform electric field applied to a periodic lattice of noninteracting fermions with a finite number of orbitals. Here the Floquet quasienergy spectrum yields the familiar Wannier-Stark (WS) ladder \cite{gluck_wannierstark_2002}. At the end of this section, we consider an alternative approach in band basis and obtain an approximate analytical solution for the WS ladder that is valid up to second order in the field.

\subsection{Lattice Hamiltonian in velocity gauge}

Our starting point is the Hamiltonian for noninteracting electrons hopping on a lattice in $D$ spatial dimensions, which can be written as
\begin{equation}
    H_0 = \sum_{\bm r, \bm r'} \sum_{a,b} \mathcal H^{ab}_{\bm r - \bm r'} c_{\bm ra}^\dag c_{\bm r'b},
\end{equation}
where $\bm r = \sum_{i=1}^D n_i \bm a_i$ (and $\bm r'$) are lattice vectors which label the cells with $n_i$ integers and $\bm a_i$ the primitive lattice vectors of the Bravais lattice. Here $a$ and $b$ are orbital indices, which includes sublattice degrees of freedom and spin. We also define the creation (annihilation) operators $c_{\bm r a}^\dag$ ($c_{\bm ra}$) which create (destroy) a fermion in cell $\bm r$ in orbital $a$. The hopping amplitude from orbital $b$ in cell $\bm r'$ to orbital $a$ in cell $\bm r$ is then given by $\mathcal H^{ab}_{\bm r - \bm r'}$ where the dependence on $\bm r - \bm r'$ reflects the translational symmetry of the system.

We now consider a uniform electric field (equivalent to taking the dipole approximation) $\bm{\mathcal E}(t)$ which can be introduced without breaking translational symmetry \cite{krieger_time_1986} in the velocity gauge via the Peierls substitution:
\begin{align}
    \mathcal H^{ab}_{\bm r} & \rightarrow \mathcal H^{ab}_{\bm r} \exp \left[ -ie \int_{\bm r_b}^{\bm r + \bm r_a} d\bm s \cdot \bm A(t) \right] \\
    & = \mathcal H^{ab}_{\bm r} e^{-ie \bm A(t) \cdot \left( \bm r + \bm r_{ab} \right)}, \label{eq:peierls}
\end{align}
with $\bm{\mathcal E}(t) = -\partial_t \bm A(t)$. Here $-e$ is the electron charge, $\bm r_a$ is the sublattice position of orbital $a$ in the unit cell, and $\bm r_{ab} = \bm r_a - \bm r_b$. The lattice Hamiltonian can then be diagonalized by Fourier transform:
\begin{equation} \label{eq:fourierc}
    c_{\bm ra} = \frac{1}{\sqrt{N}} \sum_{\bm k} e^{i \bm k \cdot \left( \bm r + \bm r_a \right)} c_{\bm ka},
\end{equation}
with $N$ the number of cells. The Hamiltonian becomes
\begin{equation} \label{eq:Hsys}
    H(t) = \sum_{\bm k} \sum_{a, b} \mathcal H^{ab}\left[ \bm k + e \bm A(t) \right] c_{\bm ka}^\dag c_{\bm kb},
\end{equation}
where the Bloch Hamiltonian is explicitly given by
\begin{equation}
    \mathcal H^{ab}(\bm k) = \sum_{\bm r} e^{-i \bm k \cdot \left( \bm r + \bm r_{ab} \right)} \, \mathcal H_{\bm r}^{ab}.
\end{equation}
Hence a uniform electric field can be introduced in the Bloch Hamiltonian by the usual minimal substitution procedure on the crystal momentum $\bm k \rightarrow \bm k + e \bm A(t)$ since the velocity gauge conserves the translational symmetry of the crystal. Note that we have retained information on the intracell positions in Eq.\ \eqref{eq:peierls} and Eq.\ \eqref{eq:fourierc} to properly account for the electrostatic potential \cite{simon_contrasting_2020}. In the (instantaneous) band basis, this choice corresponds to the \emph{periodic gauge} where the total Bloch wave function is defined on a torus: $\left| \psi_{s,\bm k+\bm g} \right> = \left| \psi_{s\bm k} \right>$ with $s$ the band index and $\bm g$ a reciprocal lattice vector \cite{vanderbilt_berry_2018}.

If we want to treat the external field perturbatively in velocity gauge, one usually considers a uniform field that oscillates with frequency $\omega$,
\begin{equation}
    \bm A(t) = \frac{|\bm{\mathcal E}|}{2i\omega} \left( \bm \epsilon e^{-i\omega t} - \bm \epsilon^* e^{i \omega t} \right),
\end{equation}
with complex polarization vector $\bm \epsilon$. Then $|\bm A(t)| \leq \bm{\mathcal E}|/\omega$ is bounded and the static limit is obtained by taking $\omega$ to zero at the end of the calculation. However, here we are interested in the nonperturbative treatment of a dc uniform electric field. Hence, we take
\begin{equation}
    \bm A(t) = -t \bm{\mathcal E},
\end{equation}
such that the Bloch Hamiltonian \cite{wannier_elements_1959}
\begin{equation}
     \mathcal H (\bm k, t) = \mathcal H \left[ \bm k(t) \right] = \mathcal H \left( \bm k - e \bm{\mathcal E} t \right).
\end{equation}
We now introduce the notion of a \emph{commensurate} electric field, which lies parallel to a reciprocal lattice vector, or equivalently perpendicular to a lattice vector,
\begin{equation}
    \bm{\mathcal E} = \mathcal E \bm g / g,
\end{equation}
where $\mathcal E >0$ and $\bm g$ is a nonzero reciprocal lattice vector with $g = |\bm g|$. For a commensurate field, we thus have
\begin{equation}
    \mathcal H(\bm k, t + T) = \mathcal H(\bm k - \bm g, t) = U_{-\bm g} \mathcal H(\bm k, t) U_{-\bm g}^\dag,
\end{equation}
with Bloch period $T = g / e \mathcal E$ and
\begin{equation} \label{eq:Hperiod}
    U_{\bm g}^{ab} = e^{-i \bm g \cdot \bm r_a} \delta_{ab},
\end{equation}
where $a,b$ denote orbital indices, and we used that $\bm g \cdot \bm r \in 2\pi \mathds Z$. For a commensurate electric field, the Bloch Hamiltonian is thus periodic in time up to a diagonal unitary transformation.

\subsection{Floquet ansatz}

The time periodicity of the Bloch Hamiltonian allows us to solve the time-dependent Schr\"odinger equation, 
\begin{equation} \label{eq:schrodinger}
    i \partial_t \left| \Phi_{\bm k}(t) \right> = \mathcal H(\bm k, t) \left| \Phi_{\bm k}(t) \right>,
\end{equation}
with a modified Floquet \emph{ansatz}
\begin{equation} \label{eq:floquetansatz}
    \Phi_{\bm ka}(t) = e^{-i \varepsilon_{\bm k} t} \sum_{n \in \mathds Z} e^{i ( n + \lambda_a ) \Omega t} \phi_{\bm ka,n},
\end{equation}
where $\varepsilon_{\bm k}$ is the quasienergy, $\Omega = 2 \pi / T$ is the Floquet frequency, and we defined $\lambda_a = \bm g \cdot \bm r_a / 2\pi$ such that
\begin{equation} \label{eq:floquet2}
    \Phi_{\bm ka}(t + T) 
    = e^{-i \varepsilon_{\bm k} T} e^{i \bm g \cdot \bm r_a} \Phi_{\bm ka}(t),
\end{equation}
which undoes the unitary in Eq.\ \eqref{eq:Hperiod} and makes Eq.\ \eqref{eq:schrodinger} invariant under $t \mapsto t + T$ even for systems with sublattice structure. Substituting the \emph{ansatz} from Eq.\ \eqref{eq:floquetansatz} into the Schr\"odinger equation, yields
\begin{equation}
    \left[ \varepsilon_{\bm k} - ( m + \lambda_a ) \Omega \right] \phi_{\bm ka,m} = \sum_{n \in \mathds Z} \sum_b \mathcal H_{mn}^{ab} \phi_{\bm kb,n},
\end{equation}
with $\mathcal H_{mn}^{ab} = \mathcal H_{m-n}^{ab}$ where
\begin{align} \label{eq:HF}
    \mathcal H_{m-n}^{ab} & = \frac{\Omega}{2 \pi} \int_0^{2\pi/\Omega} dt \, e^{-i ( m - n + \lambda_a - \lambda_b ) \Omega t} \mathcal H^{ab}(\bm k, t) \\
    \begin{split}
        & = \sum_{\bm r} e^{-i \bm k \cdot \left( \bm r + \bm r_{ab} \right)} \, \mathcal H_{\bm r}^{ab} \\
        & \qquad \times \frac{\Omega}{2 \pi} \int_0^{2\pi / \Omega} dt \, e^{-i \left( m - n - \bm g \cdot \bm r / 2 \pi \right) \Omega t}
    \end{split} \\
    & = \sum_{\bm r} e^{-i \bm k \cdot \left( \bm r + \bm r_{ab} \right)} \, \mathcal H_{\bm r}^{ab} \delta_{m-n,\bm g \cdot \bm r /2\pi},
\end{align}
where $\bm g \cdot \bm r / 2\pi$ is an integer by definition. The Floquet Hamiltonian is then defined as
\begin{equation} \label{eq:Hfloquet}
    \left[ H_F(\bm k) \right]_{mn}^{ab} = \Omega \left( m + \lambda_a \right) \delta_{mn} \delta_{ab} +  \mathcal H_{m-n}^{ab}(\bm k),
\end{equation}
with $H_F(\bm k) \left| \phi_{\bm k} \right> = \varepsilon_{\bm k} \left| \phi_{\bm k} \right>$. Here the first term is interpreted as the potential energy of a charge $-e$ located at site $\bm r + \bm r_a$. Indeed, we have $ \Omega( m + \lambda_a ) = e\bm{\mathcal E} \cdot \left( \bm r + \bm r_a \right)$ for $\bm g \cdot \bm r = 2\pi m$ where $\Omega \lambda_a$ is a field-induced sublattice potential. The velocity operator in Floquet representation is then given by
\begin{equation}
    \begin{aligned}
        \left( \nabla_{\bm k} H_F \right)_{mn}^{ab} & = - i \sum_{\bm r} \left( \bm r + \bm r_{ab} \right) e^{-i \bm k \cdot \left( \bm r + \bm r_{ab} \right)} \\
        & \qquad \times \mathcal H_{\bm r}^{ab} \delta_{m - n, \bm g \cdot \bm r /2\pi},
    \end{aligned}
\end{equation}
which contains both intercell and intracell contributions in periodic gauge. Note that the Floquet quasienergies correspond to the WS ladder \cite{wannier_wave_1960,kane_zener_1960,shockley_stark_1972}. Moreover, the quasienergy is \emph{flat} in the momentum direction parallel to the electric field. This can be understood as the WS localization of Bloch states in a static electric field \cite{emin_existence_1987}. Indeed, we notice that
\begin{equation}
    \mathcal H \left[ \bm k(t) \right] = \mathcal H \left( \bm k_\perp - \frac{t - t_0}{T} \, \bm g \right),
\end{equation}
with $\bm k_\perp = \bm k - \left( \bm k \cdot \bm g / g^2 \right) \bm g$ such that $\bm k_\perp \cdot \bm g = 0$ and $t_0 = T \bm k \cdot \bm g / g^2$. Hence the momentum parallel to the electric field can always be removed by shifting the origin of time, i.e., by performing a gauge transformation of the vector potential, such that $\varepsilon_{\bm k} = \varepsilon_{\bm k_\perp}$. Hence, the time-independent problem in length gauge in $D$ spatial dimensions is effectively mapped to a time-dependent problem in velocity gauge in $D-1$ spatial dimensions \cite{gomez-leon_floquet-bloch_2013}. Equivalently, we have mapped the physical direction of the electric field to a synthetic Floquet dimension which is possible only if the electric field lies along a reciprocal lattice vector.

\subsection{Band projection, hybrid Wannier basis, and interband corrections}

\begin{figure}
    \centering
    \includegraphics[width=.8\linewidth]{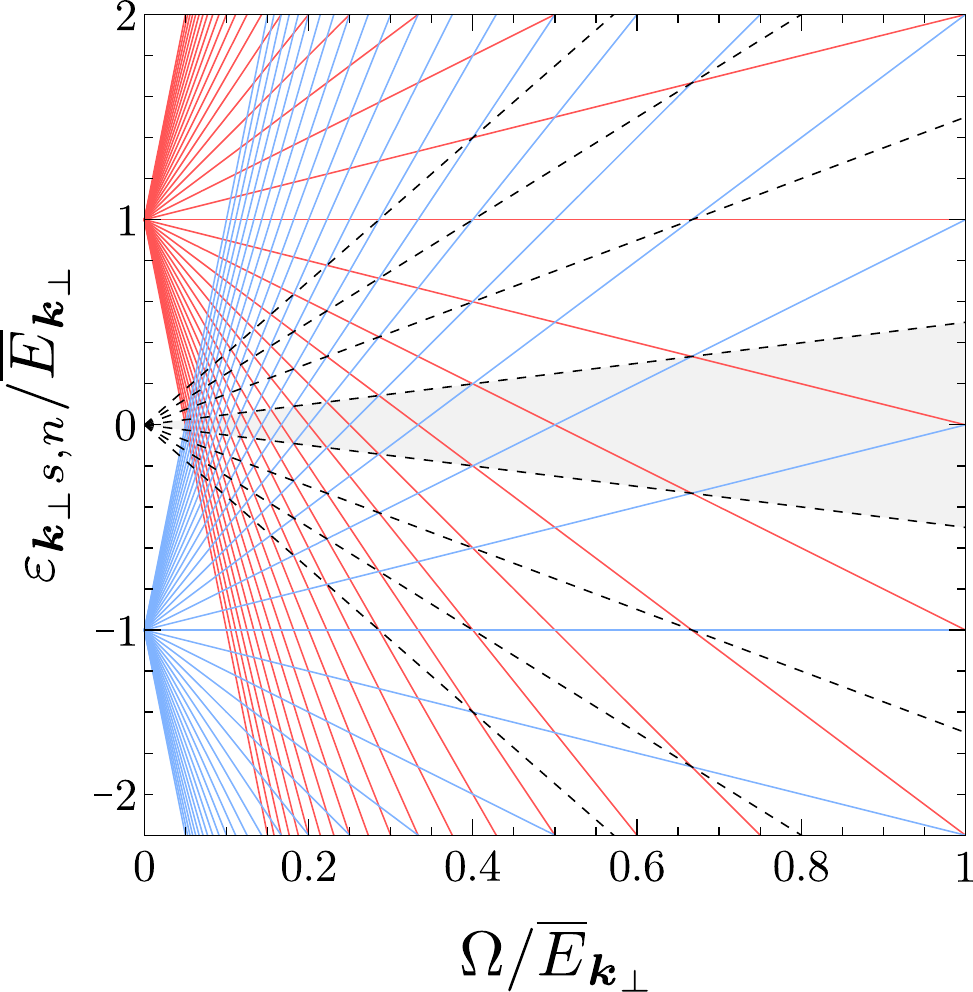}
    \caption{\textbf{Band-projected WS ladder} for a two-band model with energy bands $\pm E_{\bm k}$ shown as the red and blue lines, respectively, and where $\bm{\mathcal E} \cdot \bm{\mathcal A}_{\bm ks} = 0$. The first Floquet zone (FZ) is shown in gray and the dashed lines mark the boundaries of the first few zones.}
    \label{fig:bandprojection}
\end{figure}

While the orbital basis is best suited to study the strong-field limit, e.g., with the Magnus expansion, a good approximation of the WS ladder in the weak-field limit can be obtained in band basis (see Appendix \ref{app:bandbasis}) by band projection. In a band-projected theory, one requires that the band $E_{\bm k s}$ under consideration is energetically isolated from other bands and that the interband terms $e \bm{\mathcal E} \cdot \bm{\mathcal A}_{ss'}$ are small, where $\bm{\mathcal A}_{ss'}$ (for $s' \neq s$) is the interband Berry connection \cite{wannier_elements_1959,avron_instability_1977,krieger_time_1986,phong_quantum_2023}. In this case, time evolution is treated adiabatically and the band index remains a good quantum number. The wave function in velocity gauge is then approximately given by the instantaneous eigenstate times a phase modulation, 
\begin{equation}
    \left| \Phi_{\bm ks}(t) \right> = a_{\bm ks}(t) \left| u_{\bm ks}(t) \right>,
\end{equation}
with
\begin{equation}
    a_{\bm k s}(t) = \exp \left\{ - i \int_{0}^t dt' \left[ E_{\bm ks}(t') + e \bm{\mathcal E} \cdot \bm{\mathcal A}_{\bm ks}(t') \right] \right\},
\end{equation}
where we put $a_{\bm ks}(0) = 1$ and where $\bm{\mathcal A}_{\bm ks}(t)$ is the instantaneous intraband Berry connection for band $s$. For a commensurate electric field, the wave function satisfies Eq.\ \eqref{eq:floquet2} and thus $a_{\bm ks}(t+T) = e^{-i\varepsilon_{\bm ks} T} a_{\bm ks}(t)$ with
\begin{equation} \label{eq:bandprojection}
    \varepsilon_{\bm k_\perp s,n} = \overline E_{\bm k_\perp s} + n \Omega + e \bm{\mathcal E} \cdot \overline{\bm{\mathcal A}}_{\bm k_\perp s},
\end{equation}
where
\begin{align}
    \overline E_{\bm k_\perp s} & = \frac{\Omega}{2\pi} \int_0^{2\pi/\Omega} dt \, E_s ( \bm k - e \bm{\mathcal E} t ), \\
    \overline{\bm{\mathcal A}}_{\bm k_\perp s} & = \frac{\Omega}{2\pi} \int_0^{2\pi/\Omega} dt \, \bm{\mathcal A}_s ( \bm k - e \bm{\mathcal E} t ), \label{eq:berryphase}
\end{align}
are respectively the average band energy and the Berry phase along the field which is also called the Zak phase \cite{zak_berrys_1989}. The latter gives the polarization, i.e., the Wannier center, along the direction of the field \cite{wannier_wave_1960}. This is illustrated in Fig.\ \ref{fig:bandprojection} for the case of two bands. Here we work in a smooth periodic gauge along the integration path such that Eq.\ \eqref{eq:berryphase} is well defined. This gauge always exists regardless of the Chern number of the band. While the Chern number is an obstruction to a smooth gauge in the \emph{entire} BZ, one can always construct a smooth and periodic gauge along one compact direction \cite{kohn_analytic_1959,vanderbilt_berry_2018}. Indeed we can always shift the obstruction away from the path $\{ \bm k_\perp + s \bm g | 0 \leq s < 1 \}$ via a gauge transformation of the cell-periodic Bloch functions. Moreover, under a gauge transformation that preserves periodic gauge, Eq.\ \eqref{eq:bandprojection} is invariant modulo $\Omega \mathds Z$ since
\begin{align}
    e \bm{\mathcal E} \cdot \int_0^T dt \, \nabla_{\bm k} \varphi[\bm k(t)] & = -\int_0^T dt \, \partial_t \varphi( \bm k - e \bm{\mathcal E} t ) \\
    & = \varphi(\bm k) - \varphi(\bm k - \bm g) \in 2\pi \mathds Z.
\end{align}
However, a different choice of sublattice positions will correspond to a different periodic gauge for the cell-periodic Bloch functions. Indeed, the intracell coordinates enter the band-projected theory only via the intraband Berry connection. Similarly, the sublattice positions also enter explicitly in the Floquet theory [$\lambda_a$ in Eq.\ \eqref{eq:Hfloquet}]. This is not too surprising as a shift of the spatial origin changes the energy in the presence of a uniform electric field. In this work, we always use a periodic gauge where we choose the spatial origin such that the intraband Berry connection is traceless ($\sum_s \bm{\mathcal A}_{\bm ks} = \bm 0$). An equivalent result for the band-projected WS ladder can also be obtained in length gauge \cite{wannier_wave_1960,avron_instability_1977,phong_quantum_2023} or via the Bohr-Sommerfeld quantization rule for Bloch oscillations \cite{xiao_berry_2010,kim_surface_2016}.

The band-projected result in Eq.\ \eqref{eq:bandprojection} has a simple interpretation in terms of a hybrid momentum space and Floquet-Wannier representation \cite{gluck_wannierstark_2002}. The first term gives the on-site energy in the hybrid Wannier basis along the field direction. The on-site term is the only nonzero term since the momentum parallel to the field can be gauged away. This is the WS localization. Obstructions to a global smooth gauge do not preclude WS localization since it only entails states that are exponentially localized along the field, which is always possible as discussed above. In this representation, we see that both the position $n$ along the field and $\bm k_\perp$ are good quantum numbers. Indeed, the electrostatic energy associated with cell $\bm r$ is given by $e\bm{\mathcal E} \cdot ( \bm r + \overline{\bm r}_{\bm k_\perp s}) = n \Omega + e \bm{\mathcal E} \cdot \overline{\bm{\mathcal A}}_{\bm k_\perp s}$ with $\bm g \cdot \bm r = 2\pi n$.

One can go beyond band projection and include interband corrections to the WS ladder by solving the dynamics in band basis up to second order in the interband matrix elements. The result is derived in Appendix \ref{app:bandbasis} and reads
\begin{equation}
    \varepsilon_{\bm k_\perp s,n} = \overline{E}_{\bm k_\perp s} + n \Omega + e \mathcal E_i \left( \overline{\mathcal A}_{\bm k_\perp s}^i + e \mathcal E_j \overline \chi^{ij}_{\bm k_\perp s} \right),
\end{equation}
where summation over repeated indices is implied and
\begin{equation} \label{eq:chi}
    \chi_{\bm ks}^{ij}(t) = \sum_{s' \neq s} \frac{\mathcal A_{ss'}^i(\bm k,t) \mathcal A_{s's}^j(\bm k,t)}{E_{\bm ks}(t) - E_{\bm ks'}(t)},
\end{equation}
with $\mathcal A_{ss'}^i(\bm k,t)$ the instantaneous interband Berry connection and where $\overline \chi^{ij}_{\bm k_\perp s}$ is the (static) electric susceptibility. The lowest-order interband correction thus gives a field-dependent shift of the polarization. To our knowledge, the electric susceptibility of Bloch bands was first discussed in length gauge in a 1955 paper by E.\ O.\ Kane \cite{kane_zener_1960} and has recently been recast in the framework of quantum geometry \cite{gao_field_2014} where it was shown to give rise to an intrinsic nonlinear Hall effect when both spatial inversion and time-reversal symmetry are broken. If moreover the system conserves their combination, the dominant intraband nonlinear Hall effect is absent because the Berry curvature vanishes. Recently, this effect was observed in thin films of topological insulators with antiferromagnetic order \cite{wang_quantum-metric-induced_2023,gao_quantum_2023}. Since the susceptibility correction only shifts the polarization, energy gaps in the WS ladder due to interband transitions only arise at higher orders in the field. These results are corroborated by a numerical diagonalization of the Floquet Hamiltonian in the orbital basis, as we will demonstrate in Section \ref{sec:examples}.

\section{Floquet-Keldysh transport theory} \label{sec:transport}

In this section, we use Floquet-Keldysh theory to calculate the steady-state current by coupling the system to a featureless ideal bath. We begin by defining the steady-state current
\begin{align}
    \bm j(t) &= -e\sum_{a,b}\int_{\bm k}{\bm v}^{ab}_{\bm k}(t) \langle c^\dagger_{{\bm k}a}(t)c_{\bm kb}(t) \rangle \\
    & \equiv ie\int_{\bm k} {\rm Tr}\left[{\bm v}_{\bm k}(t)G^<({\bm k};t,t)\right],
\end{align}
where $a$, $b$ are orbital indices, $\bm v_{\bm k}^{ab}(t) = \nabla_{\bm k} \mathcal H^{ab} (\bm k,t)$ is the instantaneous velocity operator, and $\int_{\bm k} \equiv \int_\text{BZ} d^D{\bm k} / (2\pi)^D$ with $D$ the number of spatial dimensions. Here we defined the fermionic steady-state lesser Green's function $[G^<({\bm k};t,t')]_{ab} = i \langle c^\dagger_{\bm kb}(t')c_{\bm k a}(t) \rangle$ in the orbital basis. Since we are interested in the range of validity of the band-projected theory, we also define the intraband and interband contributions to the current:
\begin{align}
    \bm j_\text{intra}(t) & = \sum_s \bm j_{ss}(t), \\
    \bm j_\text{inter}(t) & = \sum_{\substack{s,s' \\  s \neq s'}} \bm j_{ss'}(t),
\end{align}
where $s,s'$ are band indices of the undriven system and the band-resolved currents are given by
\begin{align}
    \bm j_{ss'}(t) & = -e\int_{\bm k} \bm v^{ss'}_{\bm k}(t) \langle c_{\bm ks}^\dag(t) c_{\bm ks'}(t) \rangle \label{eq:jband1} \\
    & = ie \int_{\bm k} \text{Tr} \left[ P_{\bm ks}(t) \bm v_{\bm k}(t) P_{\bm ks'}(t) G^<(\bm k; t, t) \right], \label{eq:jband2}
\end{align}
where we defined the instantaneous band projectors $P_{\bm ks}(t) = \left| u_{\bm ks}(t) \right> \left< u_{\bm ks}(t) \right|$ and where the total current $\bm j(t) = \bm j_\text{intra}(t) + \bm j_\text{inter}(t)$. Here, we prefer Eq.\ \eqref{eq:jband2} over Eq.\ \eqref{eq:jband1} since it is more convenient to calculate the lesser Green's function in the orbital basis and use band projectors than it is to calculate the lesser Green's function in the band basis. Our next goal is to rewrite the current in the Floquet basis. To this end, we first define the Green's function in frequency space,
\begin{equation} \label{eq:Gft}
    G^<(\bm k; t, t') = \int_{\omega} \int_{\omega'} \, G^<(\bm{k}; \omega, \omega') \, e^{i\omega t} e^{- i \omega' t'},
\end{equation}
where $\int_\omega \equiv \int_{-\infty}^{+\infty} d\omega/ 2\pi$. The modified Floquet \emph{ansatz} of Eq.\ \eqref{eq:floquetansatz} implies that
\begin{equation} \label{eq:foobar}
    [ G^<(\bm k; t+T, t'+T) ]_{ab} = e^{2 \pi i ( \lambda_b - \lambda_a )} [ G^<(\bm k; t, t') ]_{ab},
\end{equation}
and therefore we must have $\omega - \omega' = ( m + \lambda_b - \lambda_a ) \Omega$ with $m \in \mathds Z$ in Eq.\ \eqref{eq:Gft}. Hence,
\begin{equation}
    \begin{aligned}
        & [ G^<(\bm k; t, t') ]_{ab} = \sum_{m,n} \int_0^\Omega \frac{d\omega}{2\pi} \int_0^\Omega \frac{d\omega'}{2\pi} \\
        & \times ( G^< )_{mn}^{ab}(\bm{k}; \omega, \omega') e^{i [ \omega - ( m + \lambda_a ) \Omega ] t} e^{-i [ \omega' - ( n + \lambda_b ) \Omega ] t'},
    \end{aligned}
\end{equation}
where we defined
\begin{align}
    & ( G^< )_{mn}^{ab}(\bm{k}; \omega, \omega') \notag \\
    & \qquad \equiv G^<[\bm{k}; \omega - ( m + \lambda_a ) \Omega, \omega' - ( n + \lambda_b ) \Omega] \\
    & \qquad = 2 \pi \delta( \omega - \omega') ( G^< )_{mn}^{ab}(\bm{k}, \omega ),
\end{align}
with $m$ and $n$ Floquet indices and where the last line follows from Eq.\ \eqref{eq:foobar}. We can then write the equal-time lesser Green's function as \cite{chinzei_disorder_2020}
\begin{equation}
    [ G^<({\bm k};t,t) ]_{ab} = \sum_m \int_\omega ( G^< )_{m0}^{ab}({\bm k},\omega) e^{-i ( m + \lambda_a - \lambda_b ) \Omega t},
\end{equation}
where we used $G_{mn}^<(\bm k,\omega) = G_{m-n,0}^<(\bm k,\omega - n\Omega)$. With this result, the current becomes
\begin{equation}
    \bm j(t) = ie \sum_{m,a,b} \int_{\bm k,\omega} \bm v_{\bm k}^{ab}(t) ( G^< )_{m0}^{ab}(\bm k, \omega) e^{-i ( m + \lambda_a - \lambda_b ) \Omega t},
\end{equation}
with the $n$th frequency component
\begin{align}
    \bm j^{(n)} & = \frac{\Omega}{2\pi} \int_0^{2\pi/\Omega} dt \, \bm j(t) \, e^{in\Omega t} \\
    & = ie \sum_m \int_{\bm k,\omega} \text{Tr}\left[\left( \nabla_{\bm k} H_F \right)_{nm} G^<_{m0}({\bm k},\omega)\right], \label{eq:jn}
\end{align}
where we used the definition of the Floquet Hamiltonian in Eq.\ \eqref{eq:HF}. However, for a static field, the \textit{steady state} is static and the ac response vanishes. This follows also from the fact that a static electric field can be treated in length gauge with a scalar potential giving a time-independent problem with a static steady state. Hence, even though the semiclassical trajectories are periodic \cite{bloch_uber_1929,phong_quantum_2023}, there is no net ac response at long times compared to the relaxation rate $1/\Gamma$. Indeed, in the steady state the density matrix of the WS ladder is completely diagonal such that interladder coherences are absent resulting in a purely dc response. In order to obtain an ac response, one requires both a static and oscillating component of the electric field \cite{tsu_superlattice_2005,de_beule_berry_2023}. Since we consider a static field ($\partial_t \bm{\mathcal E} = 0$) the \emph{steady state} current is therefore static [$\bm j(t) = \bm j$] regardless of $\Omega / \Gamma $. We show this explicitly in Appendix \ref{app:1dchain} for a simple chain in one spatial dimension, for which one can obtain a closed-form expression for the current. 

The band-resolved currents become
\begin{equation} \label{eq:jband_dc}
    \begin{aligned}
        & \bm j_{ss'} = ie \sum_m \sum_{l,l'} \int_{\bm k,\omega} \\
        & \, \times \text{Tr} \left[ \left( P_{Fs} \right)_{0l} \left( \nabla_{\bm k} H_F \right)_{ll'} \left( P_{Fs'} \right)_{l'm} G^<_{m0}({\bm k},\omega)\right],
    \end{aligned}
\end{equation}
with projectors in the Floquet basis
\begin{equation}
    \left( P_{Fs} \right)_{mn}^{ab} = \frac{\Omega}{2\pi} \int_0^{2\pi / \Omega} dt \, P_{\bm ks}^{ab}(t) e^{-i ( m - n +\lambda_a - \lambda_b )\Omega t},
\end{equation}
where we suppressed the momentum index on the left-hand side.

So far, we have assumed that the system reaches a steady state. To achieve this, we couple the system to a bath and calculate the lesser Green's function $G^<$ within the Keldysh formalism \cite{morimoto_topological_2016}. We note that this calculation can equally well be performed in first quantization \cite{matsyshyn_fermi-dirac_2023}. The details of the bath and the Keldysh calculation are given in Appendix \ref{app:keldysh} and yield
\begin{equation}
    G^< = G^R \Sigma^< G^A,
\end{equation}
where $G^R$ and $G^A = \left( G^R \right)^\dag$ are the retarded and advanced Green's functions, and
\begin{equation}
    \Sigma^< = \frac{\Sigma^A - \Sigma^R + \Sigma^K}{2},
\end{equation}
is the lesser self energy that takes into account the coupling of the system to the bath, where $\Sigma^R$, $\Sigma^A$, and $\Sigma^K$ are the retarded, advanced, and Keldysh self energies, respectively. We now introduce a simple model for the bath that retains translational symmetry by coupling each unit cell to its own bath in thermal equilibrium \cite{morimoto_topological_2016, kamenev_field_2011}. In this case, the Green's functions,
\begin{equation}
    G^{R/A}_{mn}(\bm k,\omega) = \left[ \omega \mathds 1 - H_F(\bm k) - \Sigma^{R/A} \right]_{mn}^{-1},
\end{equation}
in the Floquet basis. Moreover, for simplicity we consider an \emph{ideal bath} by taking a constant density of states $\rho_0$ (wide-band limit) and a constant coupling $\lambda$ between the system and the bath \cite{jauho_time-dependent_1994,matsyshyn_fermi-dirac_2023}. For an ideal bath, the self energies in the Floquet basis take the form
\begin{align}
    (\Sigma_+)_{mn}^{ab} & =
    \begin{pmatrix}
    (\Sigma_+^K)_{mn}^{ab} & (\Sigma_+^R)_{mn}^{ab} \\ (\Sigma_+^A)_{mn}^{ab} & 0 \end{pmatrix} \\
    & = -i \Gamma \delta_{mn} \delta_{ab}
    \begin{pmatrix}
    1+2f_+^0[ \omega - ( m + \lambda_a ) \Omega ] & \frac{1}{2} \\
    -\frac{1}{2} & 0 \end{pmatrix}, \label{eq:selfenergy-bose}
\end{align}
for an ideal bosonic bath (e.g.\ phonons), or
\begin{align}
    (\Sigma_-)_{mn}^{ab} & =
    \begin{pmatrix}
    (\Sigma_-^R)_{mn}^{ab} & (\Sigma_-^K)_{mn}^{ab}\\ 0 & (\Sigma_-^A)_{mn}^{ab} \end{pmatrix} \\
    & = -i \Gamma \delta_{mn} \delta_{ab}
    \begin{pmatrix}
    \frac{1}{2} & 1-2f_-^0[ \omega - ( m + \lambda_a ) \Omega ] \\
    0 & -\frac{1}{2} \end{pmatrix}, \label{eq:selfenergy-fermi}
\end{align}
for an ideal fermionic bath (i.e.\ leads attached to each site). Here we defined $\Gamma = \lambda^2 \rho_0$ whose role could alternatively have been played by a disorder-averaged scattering rate $\Gamma = 1/\tau$ \cite{coleman_introduction_2015} and $f^0_\pm(\omega) = 1 / \left[ e^{\beta (\omega - \mu)} \mp 1 \right]$ is the equilibrium distribution for fermions and bosons, respectively, with inverse temperature $\beta$ and chemical potential $\mu$. The diagonal components in Eq.\ \eqref{eq:selfenergy-fermi} and Eq.\ \eqref{eq:selfenergy-bose} give the retarded and advanced self energies induced by the bath and the Keldysh component reflects the fluctuation-dissipation theorem for the bath: $\Sigma^K = ( 1 \pm 2f_\pm^0 ) ( \Sigma^R - \Sigma^A )$ \cite{jauho_introduction_2006, kamenev_field_2011, eissing_functional_2016}. Putting everything together, we obtain
\begin{equation} \label{eq:lesser_selfenergy}
    ( \Sigma_\pm^< )_{mn}^{ab} = \mp i \Gamma \delta_{mn} \delta_{ab} f_\pm^0[ \omega - ( m + \lambda_a ) \Omega ].
\end{equation}
We note that Eq.\ \eqref{eq:lesser_selfenergy} coincides for an ideal bosonic or fermionic bath in the low-temperature limit, since $f_+^0(\omega) = - f_-^0(\omega)$ when $\beta |\omega - \mu|$ becomes large.  Combining Eq.\ \eqref{eq:jband_dc} and Eq.\ \eqref{eq:lesser_selfenergy}, we finally obtain for the band-resolved currents,
\begin{widetext}
\begin{equation} \label{eq:jband_final}
    \bm j_{ss'} = \pm e\Gamma \sum_{m,m'} \sum_{l,l'} \int_{\omega,{\bm k}} \text{Tr} \left[  \left( P_{Fs} \right)_{0l} \left( \nabla_{\bm k} H_F \right)_{ll'} \left( P_{Fs'} \right)_{l'm'} G^R_{m'm}(\omega,\bm k) \mathcal F^0_\pm (\omega - m \Omega) G_{m0}^A(\omega,\bm k) \right],
\end{equation}
\end{widetext}
with $[ \mathcal F^0_\pm (\omega - m \Omega) ]_{ab} = \delta_{ab} f^0_\pm [ \omega - ( m + \lambda_a ) \Omega]$. While the frequency integral can be done analytically by using the single-particle Lehmann representation of the Floquet Green's functions $G^{R/A}$, we find in practice that it is faster to perform the frequency integral numerically, where we ensure convergence by taking a step size of $\Gamma/2$ which is below the energy resolution of the Green's functions. Nevertheless, the analytical results are useful in certain cases. For example, in Appendix \ref{app:integral} we conclusively show that in the zero-temperature limit, while also keeping $\beta \Gamma$ fixed such that the coupling to the bath is also taken to zero, the ideal bosonic and fermionic bath both yield the same results. For this reason, we will only consider a fermionic bath in the remainder of the paper.
\begin{figure}
    \centering
    \includegraphics[width=.85\linewidth]{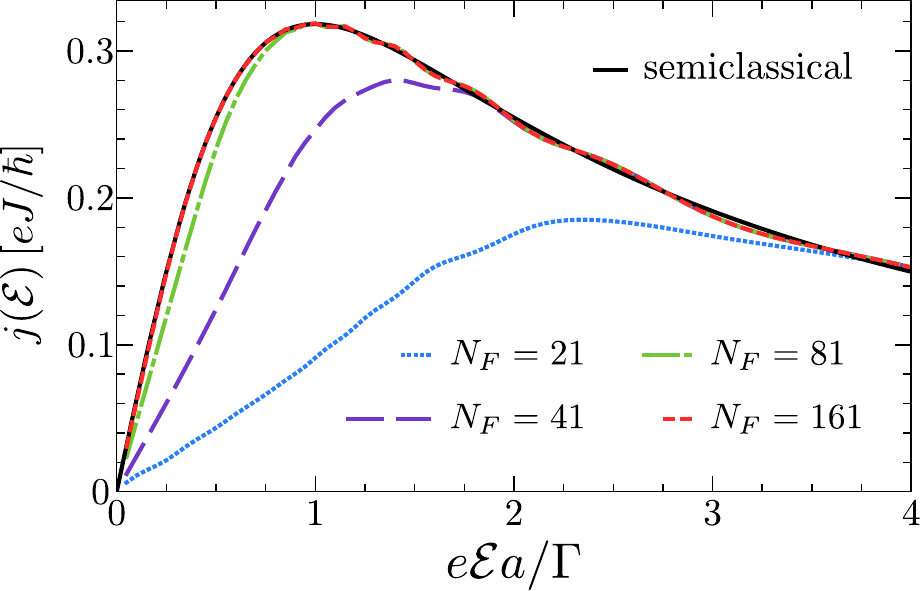}
    \caption{\textbf{Convergence of the current for the simple 1D chain coupled to an ideal fermionic bath.} Steady-state current for the simple chain at half filling with $\beta J = 50$ and $\Gamma/J = 0.1$ for different number of Floquet sidebands as indicated. We only show the current for $\bm{\mathcal E} = \mathcal E \bm{\hat x}$ with $\mathcal E > 0$ since inversion symmetry yields $j(-\mathcal E) = -j(\mathcal E)$. The solid line gives the semiclassical result [see Eq.\ \eqref{eq:j_chain_sc}]. Here we do not show the exact result (see Appendix \ref{app:integral}) as it is indistinguishable from the converged numerical result.}
    \label{fig:j_1band}
\end{figure}

\section{Examples} \label{sec:examples}

In this section, we calculate the current in response to a static electric field up to nonperturbative order for noninteracting electrons coupled to an ideal fermionic bath with the Floquet quantum theory. We start with a one-band model in $D=1$ spatial dimensions for which the current can be obtained analytically in a closed-form expression. We then consider the effect of Zener tunneling with a two-band model in $D=1$. Finally, we consider a two-band model in $D=2$ on a honeycomb lattice, for which we closely investigate the quantum-geometric transverse response.

\subsection{Examples in $D=1$}

\subsubsection{Simple chain}

We first consider a chain with a single orbital per site in $D=1$ spatial dimensions with lattice constant $a$ and nearest-neighbor hopping $J$. In this case,
\begin{equation}
    \mathcal H(k,t) = 2 J \cos \left( ka - \Omega t \right),
\end{equation}
with $\Omega = eEa$. The quasienergies are obtained from
\begin{equation}
    \frac{\varepsilon - n \Omega}{J} \, \phi_{kn} = e^{-ika} \phi_{k,n-1} + e^{ika} \phi_{k,n+1}.
\end{equation}
If we substitute $\phi_{kn} = e^{-ikna} \varphi_n$, we find normalizable solutions for $\varepsilon = m \Omega$,
\begin{equation}
    \frac{m - n}{J / \Omega} \, \varphi_n = \varphi_{n-1} + \varphi_{n+1},
\end{equation}
which is the recurrence relation for Bessel functions. The solution is thus given by
\begin{equation}
    \phi_{kn} = e^{-ikna} J_{m-n} \left( 2J / \Omega \right),
\end{equation}
and for $\varepsilon = 0$ the wave function is given by
\begin{align}
    \Phi_k(t) & = \sum_n e^{in \left( ka - \Omega t \right)} J_n \left( 2J / \Omega \right) \\
    & = e^{2iJ \sin(ka - \Omega t) / \Omega},
\end{align}
which can also be obtained from directly integrating the Schr\"odinger equation. Moreover, the corresponding Floquet-Wannier function centered at the origin is \cite{gluck_wannierstark_2002}
\begin{align}
    W_m(t) & = \frac{a}{2\pi} \int_0^{2\pi/a} dk \, \Phi_k(t) e^{-ikma} \\
    & = e^{-i m \Omega t} J_m \left( 2J / \Omega \right),
\end{align}
whose spread is $( \Delta x )^2 = \left< x^2 \right> - \left< x \right>^2= 2 \left( J a / \Omega \right)^2$. Here we used $J_{-m}(z) = (-1)^m J_m(z)$ for real $z$ and the generating function for Bessel functions. This coincides with the period-averaged spread of the semiclassical trajectory, which for $k(0) = 0$, reads $x(t) = x(0) + ( 2 J a / \Omega ) \left[ \cos( \Omega t) - 1 \right]$. Alternatively, we can understand this in terms of a Floquet-induced quantum metric that bounds the spread of the Wannier function:
\begin{equation}
    \mathfrak g = \sum_n \left[ | \partial_k \phi_{kn} |^2 - \left( \phi_{kn}^* i \partial_k \phi_{kn} \right)^2 \right].
\end{equation}

We now consider the current $j(\mathcal E) \bm{\hat x}$ for $\bm{\mathcal E} = \mathcal E \bm{\hat x}$. Since there is only a single band in this case, the results for the simple chain are somewhat trivial. Therefore, we will use the one-band problem to demonstrate convergence as we take into account more Floquet modes in the Floquet-Keldysh calculation. This is shown in Fig.\ \ref{fig:j_1band} for an ideal fermionic bath, where we plot the current as a function of $\Omega \tau = e \mathcal E a / \Gamma$ (by fixing $\Gamma / J$ and varying $e \mathcal E a / J$) for different numbers of Floquet sidebands. Here we only consider positive values for $e \mathcal E a / J$ because 1D inversion symmetry ($x \mapsto -x$) gives $j(-\mathcal E) = -j(\mathcal E)$. We find that convergence is reached for $N_F \sim 100$ where $N_F$ is the dimension of the Floquet Hamiltonian used in the numerical calculation. In general, one requires more Floquet modes to reach convergence when $\Omega / J \ll 1$ since there are many overlapping Floquet replicas in this case. In the opposite limit, we have $\Omega / J \gg 1$ such that the Floquet replicas are well separated and few Floquet modes are required to reach convergence.

Apart from small oscillations that appear for $e \mathcal E a / \Gamma > 1$, we see that the converged current coincides almost perfectly with the semiclassical theory. This is to be expected since band projection is exact if there is only one band. Finally, we find that the oscillations are suppressed when $\Gamma$ is reduced. Contrary to the semiclassical theory, the current is thus generally a function of both $\Omega$ and $\Gamma$ separately. Using the exact closed-form expression of the current for an ideal fermionic and bosonic bath, we find that the oscillations are more pronounced in the bosonic case and vanish upon decreasing $\Gamma$ (see Appendix \ref{app:1dchain}). However, reducing $\Gamma$ increases the number of Floquet sidebands required to reach convergence, as we need smaller $\Omega$ to obtain the same $e \mathcal E a / \Gamma$.

\subsubsection{Rice-Mele chain}

\begin{figure}
    \centering
    \includegraphics[width=\linewidth]{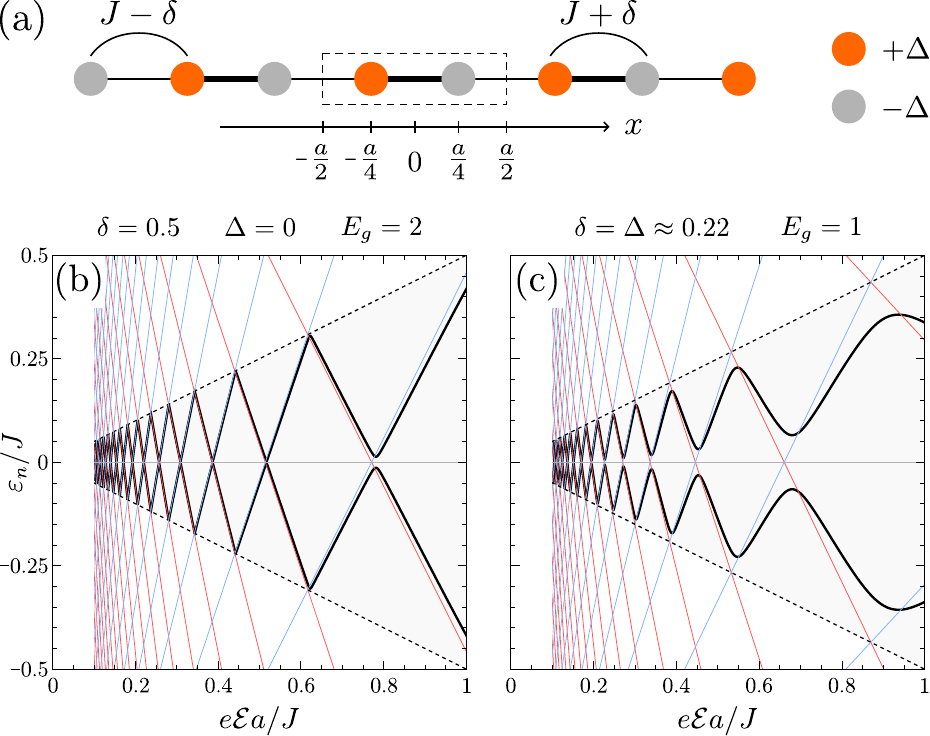}
    \caption{\textbf{WS ladder of the 1D Rice-Mele chain.} (a) Illustration of the RM model. The dashed rectangle gives the unit cell and the hopping amplitudes and on-site energies are indicated. (b) WS ladder for $E_g = 2$. (c) WS ladder for $E_g = 1$. Parameters are shown in units of $J$. The thin red and blue lines give the band-projected result for the conduction and valence band, respectively, and black lines give the quasienergy in the FZ, demarcated by dashed lines, obtained from diagonalizing the Floquet Hamiltonian numerically.}
    \label{fig:rm}
\end{figure}

In order to address the role of interband Zener transitions, we need to consider a multiband system. For simplicity, we study a two-band model,
\begin{equation} \label{eq:2band}
    \mathcal H(k) = d_0(k) \sigma_0 + \bm d(k) \cdot \bm \sigma,
\end{equation}
where $\sigma_0$ is the $2\times2$ unit matrix and $\bm \sigma = (\sigma_x, \sigma_y, \sigma_z)$ are the Pauli matrices. The energy bands are given by $E_{ks} = d_0 + sd$ with $s = \pm1$ the band index and $d=|\bm d|$. Specifically, we consider the Rice-Mele (RM) dimer chain \cite{rice_elementary_1982} illustrated in Fig.\ \ref{fig:rm}(a) with
\begin{equation}
    d_0 = 0, \qquad \bm d = 
    \begin{bmatrix}
        2 J \cos \left( \frac{ka}{2} \right) \\
        -2 \delta \sin \left( \frac{ka}{2} \right) \\
        \Delta
    \end{bmatrix},
\end{equation}
where $a$ is the lattice constant and we set $J = 1$. Here $1 \pm \delta$ is the intracell/intercell hopping amplitude, respectively, and $\Delta$ is a sublattice bias potential. We further choose sublattice positions $x_{A/B} = \pm a / 4$ which fixes the periodic gauge: $u_{k+2\pi/a,\sigma} = e^{-i2\pi x_b/a} u_{k\sigma}$ ($\sigma = A,B$) or explicitly,
\begin{equation}
    \left| u_{ks} \right> = \frac{e^{ika/4}}{\sqrt{2 ( 1 + sn_3 )}} \begin{pmatrix} n_3 + s \\ n_1 + i n_2 \end{pmatrix},
\end{equation}
with $\bm n = \bm d / d$ and which together with $d_0 = 0$ results in a symmetric quasienergy spectrum. This gauge is smooth everywhere, i.e., $\psi_k(x_a) = e^{ikx_a} u_{ka}$ is smooth and periodic, since $n_3$ never equals $-s$.

\begin{figure}
    \centering
    \includegraphics[width=\linewidth]{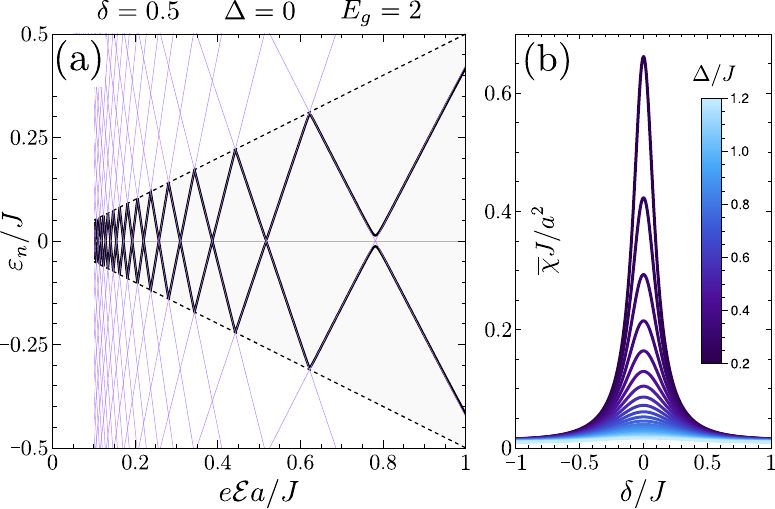}
    \caption{\textbf{Electric susceptibility.} (a) Comparison between the numerical results (black lines) and the analytical result beyond band projection (thin purple lines) up to second order in the field, which includes the polarization correction arising from interband transitions for the RM chain with the same parameters as in Fig.\ \ref{fig:rm}(b). We only show the numerical result in the FZ which is demarcated by dashed lines. (b) Electric susceptibility $\overline \chi$ for the RM chain.}
    \label{fig:rmshift}
\end{figure}

To calculate the WS ladder we send $k \mapsto k - eEt$ and construct the Floquet Hamiltonian where we gauge away the momentum with the unitary transformation
\begin{equation}
    \left( U_k \right)_{mn} = e^{ika(m\sigma_0 + \sigma_z/4)} \delta_{mn},
\end{equation}
such that the two bands in the first Floquet zone (FZ) are flat, although their splitting varies with $\Omega$. The Floquet spectrum is obtained from
\begin{align}
    & \left[ \varepsilon - ( m + \lambda_A ) \Omega - \Delta \right] \varphi_{Am} \notag \\
    & \qquad \qquad = ( 1 + \delta ) \varphi_{Bm} + ( 1 - \delta ) \varphi_{B,m-1}, \\
    & \left[ \varepsilon - ( m + \lambda_B ) \Omega + \Delta \right] \varphi_{Bm} \notag \\
    & \qquad \qquad = ( 1 + \delta ) \varphi_{Am} + ( 1 - \delta ) \varphi_{A,m+1}.
\end{align}
In the molecular dimer limit $\delta \rightarrow 1$, there is an exact solution with quasienergy
\begin{equation}
    \lim_{\delta \rightarrow 1} \varepsilon_{\pm,n} = n \Omega \pm \sqrt{4 + \left( \tfrac{\Omega}{4} - \Delta \right)^2},
\end{equation}
giving the usual Stark ladder of a dimer of length $a/2$ centered at position $x = na$. A similar result holds for $\delta \rightarrow -1$, but now the dimers are centered at $x = (n+1/2)a$. For the general case, we diagonalize $H_F$ numerically. 
Some results are shown in Fig.\ \ref{fig:rm}, where we take both a large and small energy gap $E_g$ relative to the range of field strengths that we consider. In Fig.\ \ref{fig:rm}(b), we take $E_g = 2$ such that interband coupling is weak and the dominant effect comes from the electric susceptibility, giving rise to a shift of the WS ladder that scales linearly with the field:
\begin{equation} \label{eq:WS2}
    \varepsilon_{\pm,n} = n\Omega \pm \left[ \overline d + \left( \frac{\overline{\mathcal A}}{a} + \frac{\overline{\chi}}{a^2} \, \Omega \right) \Omega\right] + \mathcal O(\Omega^3),
\end{equation}
with 
\begin{equation}
    \chi_{ks}(t) = \frac{[ \partial_k \bm n(k,t) ] \cdot [ \partial_k \bm n(k,t) ]}{8sd(k,t)},
\end{equation}
for a two-band model. In this case, $\chi$ is given by the quantum metric divided by the energy gap. This makes sense as the quantum metric determines the spread of the maximally-localized Wannier state. Increasing the spread thus results in a larger electrostatic potential difference across the support of the Wannier state and hence to an increase in the electric susceptibility. The band-projected result is recovered from the linear in $\Omega$ part of Eq.\ \eqref{eq:WS2} and shown as the thin lines in Fig.\ \ref{fig:rm}, while the full result of Eq.\ \eqref{eq:WS2} containing the lowest-order interband coupling is compared to the numerical results in Fig.\ \ref{fig:rmshift}. We see that the shift induced by the susceptibility matches perfectly to the numerical results for the parameters chosen. This matching becomes less good when we decrease the energy gap or increase the field strength, both of which increase the magnitude of interband matrix elements.
For the {RM} chain, we explicitly find
\begin{equation} 
    \overline d = \frac{\sqrt{4 + \Delta^2} E \left[ \frac{4 - 4 \delta^2}{4 + \Delta^2} \right] + \sqrt{4 \delta^2 + \Delta^2} E \left[ \frac{4\delta^2 - 4}{4\delta^2 + \Delta^2} \right] }{\pi},
\end{equation}
with $E(k) = \int_0^{\pi/2} d\theta \, \sqrt{1 - k \sin^2 \theta}$ the complete elliptic integral of the second kind, and
\begin{equation} 
    \frac{\overline{\mathcal A}}{a} = \frac{\sgn(\delta) - 1}{4} - \frac{\delta \Delta \Pi \left[ 1 - \delta^2, \frac{4 - 4\delta^2}{4 + \Delta^2} \right]}{2 \pi \sqrt{4 + \Delta^2}},
\end{equation}
in periodic gauge and our choice of sublattice positions where $\Pi(n,k) = \int_0^{\pi/2} d\theta / \left[  ( 1 - n \sin^2 \theta ) ( 1 - k \sin^2 \theta )^{1/2} \right]$ is the complete elliptic integral of the third kind. We further obtain
\begin{equation}
    \frac{\overline{\chi}}{a^2} = \frac{\left( 8 + 8 \delta^2 + \Delta^2 \right) E \left[ \frac{4 - 4 \delta^2}{4 + \Delta^2} \right] - \left( 4 \delta^2 + \Delta^2 \right) K \left[ \frac{4 - 4 \delta^2}{4 + \Delta^2} \right]}{48 \pi \sqrt{4 + \Delta^2} \left( 4 \delta^2 + \Delta^2 \right)},
\end{equation}
where $K(k) = \int_0^{\pi/2} d\theta / \sqrt{1 - k \sin^2 \theta}$ is the complete elliptic integral of the first kind. On the other hand, when the energy gap becomes sufficiently small relative to the electric field, higher-order interband Zener transitions induce gaps in the WS ladder. This can be seen in Fig.\ \ref{fig:rm}(c) for $E_g = 1$. Now there is significant band mixing and a large Zener gap opens up as the field strength increases.
\begin{figure*}
    \centering
    \includegraphics[width=.95\linewidth]{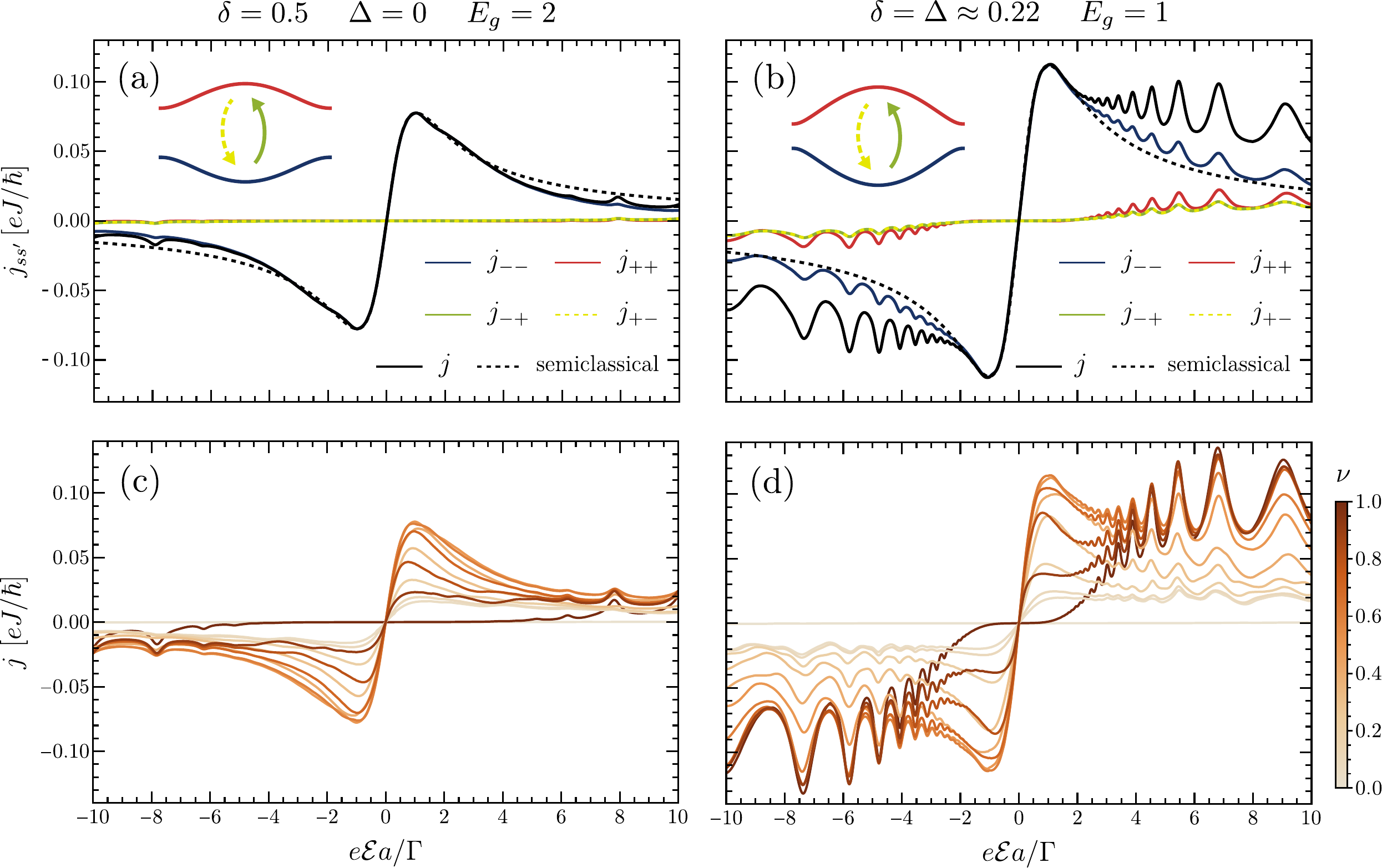}
    \caption{\textbf{Current for the RM chain.} (a-b) Band-resolved currents $j_{ss'}$ for the RM chain for two different choices of $\delta$ and $\Delta$. The corresponding energy bands are shown in the inset. Here we choose the chemical potential $\mu$ so that the lower band is half-filled. The semiclassical result is plotted as the black dashed line. (c-d) The total current for different fillings $\nu$ of the lower band. In all plots, we set $\beta=50$ and $\Gamma=0.1$ where energy is measured in units of $J$.}
    \label{fig:j_rm}
\end{figure*}

We proceed to consider the band-resolved currents for the RM chain. To this end, we need the instantaneous band projectors. In general, for a two-band model as given by Eq.\ \eqref{eq:2band}, these are given by
\begin{equation}
    P_{ks}(t) = \frac{1}{2} \left[ \sigma_0 + s \bm n(k,t) \cdot{\bm\sigma} \right].
\end{equation}
Results for the currents for the same parameters as shown in Fig.\ \ref{fig:rm}(b) and (c) for the WS ladders, are shown in Fig.\ \ref{fig:j_rm}(a) and (b), respectively. We note that the effect of the field-induced polarization shift cannot be seen in the transport calculation since in all cases considered there is a cancellation owing to time-reversal or spatial inversion symmetry. For small fields $e\mathcal E a/\Gamma \lesssim 1$, the system evolves adiabatically, leaving the current predominantly intraband. In this regime, we find perfect matching to the semiclassical result. For large fields, on the other hand, interband transitions become important, and the current deviates from the semiclassical result. We find that the peaks in the current correspond to the avoided crossings in the WS ladder (see Fig.\ \ref{fig:rm}) where the Zener transition rate is maximal. As expected, these Zener resonances in the current become larger in magnitude when the energy gap $E_g$ is small relative to the electric field. As a function of the filling $\nu$ of the bottom band, we further find that while the current for small fields is maximal near half filling, the current for large fields peaks for a \textit{fully filled} bottom band. This is shown in Fig.\ \ref{fig:j_rm}(c) and (d). We attribute this to the fact that since the Zener transitions are driven by the interband Berry connection, which is largest in magnitude near the top of the lower band, the interband current is greatest when the states near the top of the band are filled. Otherwise, the field has to first drive a significant population to the top of the band on a time scale smaller than the relaxation rate. A rough estimate of the validity of the band-projected theory in the strong-field regime can be obtained by demanding that the lowest-order interband correction is small:
\begin{equation}
    \frac{E_g^2}{W} \sim \frac{a^2 E_g}{|\mathcal A_\text{inter}|^2} \gg \Omega \gg \Gamma,
\end{equation}
where $W$ is the bandwidth. A similar bound can be found from semiclassical considerations \cite{ashcroft_solid_1976,phong_quantum_2023,de_beule_roses_2023}. Note that we have four relevant energy scales in our problem: the energy gap $E_g$, the bandwidth $W$, the electrostatic energy $e \mathcal E a$, and the system-bath coupling $\Gamma$. 

\subsection{Example in $D=2$: honeycomb lattice}

\begin{figure}
    \centering
    \includegraphics[width=\linewidth]{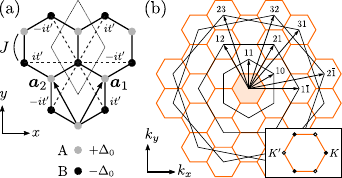}
    \caption{\textbf{Honeycomb lattice.} (a) Illustration of the honeycomb lattice. The dotted diamond is the unit cell and the nearest-neighbor (solid), next-nearest-neighbor (dashed) hopping amplitudes, and on-site energies are indicated. Here $t' = 2 \Delta_1 / (3 \sqrt{3} )$ and we only show the nnn hopping for sublattice $A$; it has an opposite sign for sublattice $B$. (b) Some commensurate field directions $\bm g = m_1 \bm g_1 + m_2 \bm g_2$ considered in this work are shown with juxtaposed coordinates $m_1m_2$ where the overline indicates a minus sign. Here the small orange hexagons are the BZs in the extended zone scheme and the large black hexagons correspond to different stars.}
    \label{fig:honeycomb}
\end{figure}

In two spatial dimensions, a general commensurate electric field is parameterized by a pair of coprime integers $(m_1,m_2)$,
\begin{equation}
    \bm{\mathcal E} = \frac{\mathcal E}{g_0} \frac{m_1 \bm g_1 + m_2 \bm g_2}{\sqrt{m_1^2 + m_2^2 - m_1 m_2}},
\end{equation}
where $\bm g_{1,2}$ are primitive reciprocal lattice vectors with length $g_0 = |\bm g_{1,2}|$ and $\bm g_i \cdot \bm a_j = 2\pi \delta_{ij}$. As a concrete example, we consider $s$ or $p_z$ electrons hopping on a honeycomb lattice with nearest-neighbor hopping amplitude $J=1$, sublattice bias $\Delta_0$, and Haldane mass $\Delta_1$ \cite{haldane_model_1988}, as illustrated in Fig.\ \ref{fig:honeycomb}. We choose primitive lattice vectors $\bm a_{1,2} = a \left( \pm 1/2, \, \sqrt{3}/2 \right)$ with $a$ the lattice constant and reciprocal vectors $\bm g_{1,2} = g_0 \left( \pm \sqrt{3}/2, \, 1/2 \right)$ with $g_0 = 4\pi / \sqrt{3} a$. The Bloch Hamiltonian in periodic gauge is then given by
\begin{equation}
\mathcal H(\bm k) = \begin{bmatrix} 
\Delta_0 + \Delta_1 \gamma(\bm k) & f \left( \bm k \right) \\
f \left( \bm k \right)^* & - \Delta_0 - \Delta_1 \gamma(\bm k) \end{bmatrix} = \bm d \cdot \bm \sigma,
\end{equation}
with $d_1 - i d_2 = f$, $d_3 = \Delta_0 + \Delta_1 \gamma$, and
\begin{align}
    f(\bm k) & = e^{i\bm k \cdot \left( \bm r_B - \bm r_A \right)} \left( 1 + e^{i \bm k \cdot \bm a_1} + e^{i \bm k \cdot \bm a_2} \right), \\
    \gamma(\bm k) & = \frac{2}{3\sqrt{3}} \left( \sin \bm k \cdot \bm a_1 - \sin \bm k \cdot \bm a_2 - \sin \bm k \cdot \bm a_3 \right),
\end{align}
with $\bm a_3 = \bm a_1 - \bm a_2$. For concreteness, we take $\bm r_{A/B} = \left( 0, \pm a / 2 \sqrt{3} \right)$ for the sublattice positions, which fixes the periodic gauge. One possibility is given by
\begin{equation}
    \left| u_{\bm k\pm} \right> = \frac{e^{-i \bm k \cdot \bm r_A}}{\sqrt{2( 1 \pm n_3 )}} \begin{pmatrix} n_3 \pm 1 \\ n_1 + i n_2 \end{pmatrix},
\end{equation}
for states with energies $\pm d$, respectively, where $d = |\bm d|$ and $\bm n = \bm d / d$. This choice gives a traceless intraband Berry connection. Note that this gauge is smooth everywhere except at the zone corners when $1 \pm n_3 = 0$ which is the case for $\sgn(\Delta_0 + \tau \Delta_1) = \mp 1$ where $\tau = \pm 1$ at the $K$ and $K'$ points, respectively [see Fig.\ \ref{fig:honeycomb}(b)]. We can try to remove this singularity with the gauge transformation
\begin{align}
    \left| \tilde u_{\bm k\pm} \right> & = \pm \frac{n_1 - i n_2}{\sqrt{n_1^2 + n_2^2}} \left| u_{\bm k\pm} \right> \label{eq:gauge} \\
    & = \frac{e^{-i \bm k \cdot \bm r_A}}{\sqrt{2( 1 \mp n_3 )}} \begin{pmatrix} n_1 - i n_2 \\ \pm 1 - n_3 \\  \end{pmatrix},
\end{align}
which is now smooth except at the zone corner where $\sgn(\Delta_0 + \tau \Delta_1) = \pm 1$. Therefore when $| \Delta_1 / \Delta_0 | < 1$, one can remove all singularities and obtain a smooth gauge by appropriately choosing either $\left| u_{\bm k\pm} \right>$ or $\left| \tilde u_{\bm k\pm} \right>$. However, for $| \Delta_1 / \Delta_0 | > 1$, the gap is inverted between $K$ and $K'$ and the gauge transformation only moves the singularity from one zone corner to the other. Moreover, taking a superposition of $\left| u_{\bm k\pm} \right>$ and $\left| \tilde u_{\bm k\pm} \right>$ allows one to place the singularity at an arbitrary point in the BZ. This obstruction to a global smooth gauge is of course due to the finite Chern number for $| \Delta_1 / \Delta_0 | > 1$, which by Stokes' theorem is given by the winding of the gauge transformation in Eq.\ \eqref{eq:gauge} around the singularity \cite{kohmoto_topological_1985}. Note that in the trivial phase if we choose to take the gauge with a singularity at both $K$ and $K'$, the net winding number vanishes. Care is taken in band basis to avoid spurious results due to these singularities.

Next, we discuss the symmetries of the honeycomb model in the presence or absence of the sublattice mass $\Delta_0$ and the Haldane mass $\Delta_1$. For $\Delta_0 = \Delta_1 = 0$, the model has time-reversal symmetry $\mathcal T$ and point group $D_{6h} = C_{6v} \times \sigma_h$ with $C_{6v} = \left< \mathcal C_{6z}, \mathcal M_x \right>$ where $\mathcal C_{6z}$ is a sixfold rotation about the $z$ axis, $\mathcal M_x: x \mapsto -x$ is the in-plane mirror with respect to the $x$ axis, and $\sigma_h$ is the mirror with respect to $z$. Since we already fixed the symmetry of the orbitals with respect to $\sigma_h$, we can restrict ourselves to $C_{6v}$. A finite sublattice potential $\Delta_0$ breaks $\mathcal C_{2z}$ symmetry, i.e., inversion symmetry when restricted to the plane, which reduces the point group to $C_{3v} = \left< \mathcal C_{3z}, \mathcal M_x \right>$. On the other hand, a finite Haldane mass $\Delta_1$ breaks $\mathcal T$ symmetry and all mirrors, allowing for a finite linear Hall response, but $\Delta_1$ conserves $\mathcal C_{2z}$ symmetry as well as combinations of mirrors and time reversal such as $\mathcal M_x \mathcal T$. This gives rise to the magnetic point group $6 \underline{m} \underline{m}$ which is also denoted as $C_{6v}(C_6)$ \cite{dresselhaus_group_2007}. In the presence of both mass terms, the magnetic point group is given by $3 \underline m$ or $C_{3v}(C_3)$.

\begin{figure}
    \centering
    \includegraphics[width=\linewidth]{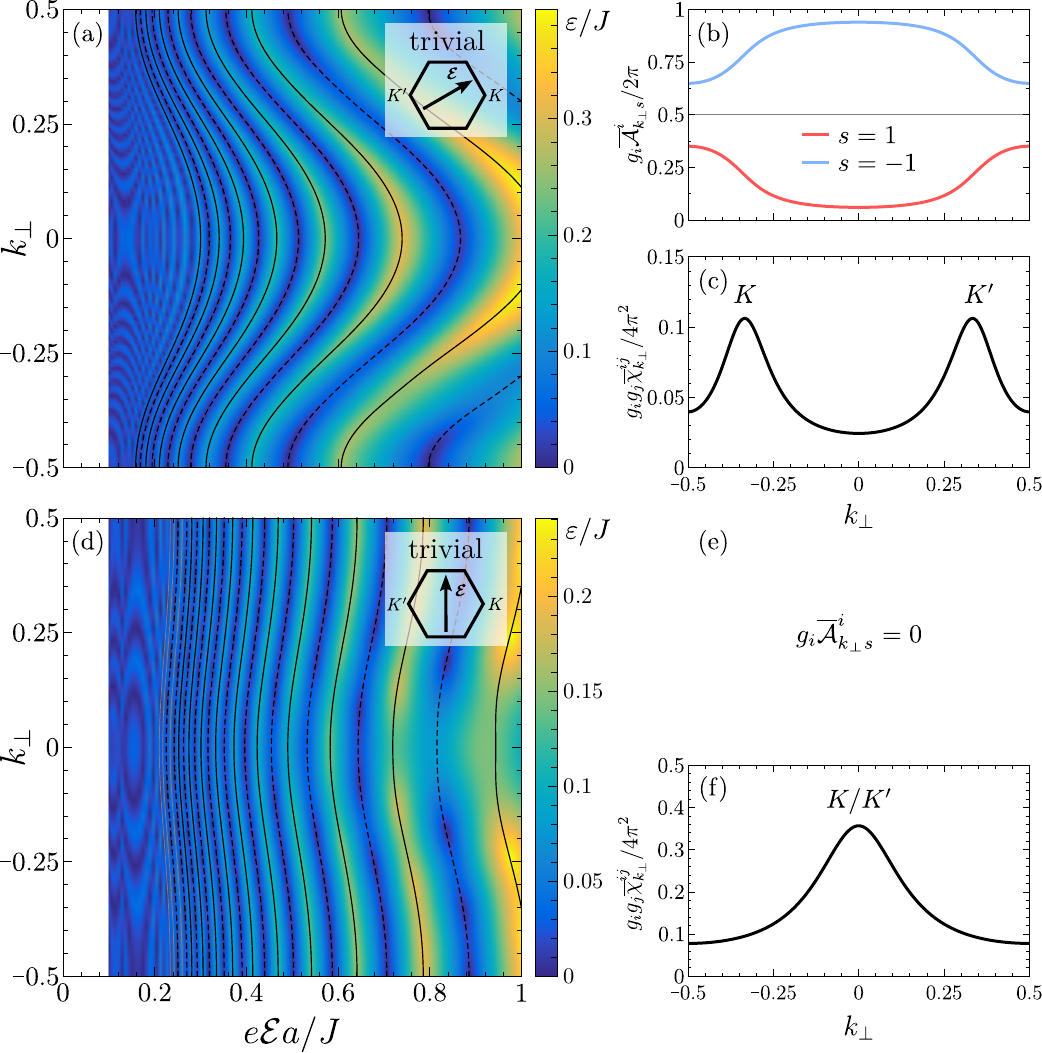}
    \caption{\textbf{WS ladder of the honeycomb lattice in the trivial phase} for $(\Delta_0,\Delta_1) = (0.5,0)$. (a) WS ladder in the FZ for $(m_1,m_2) = (1,0)$. The solid (dashed) curves give the crossings in the ladder at zero (edge of the FZ) as calculated in band basis up to 2nd order. We only show positive quasienergies since the spectrum is symmetric in traceless gauge. (b) Projected Berry phase (Wannier center) for the two bands and (c) susceptibility $\overline \chi_{k_\perp s} = s\overline \chi_{k_\perp}$ as a function of $k_\perp$ for the field in (a). (d) WS ladder, (e) Berry phase, and (f) susceptibility for $(m_1,m_2) = (1,-1)$. In this case, the Berry phase vanishes because the field lies perpendicular to a mirror line.}
    \label{fig:hc-ws-trivial}
\end{figure}

We now consider the WS ladder of the honeycomb model. To this end, we introduce the dimensionless momentum $k_\perp$ normal to the field direction $\bm g$ such that a general momentum can be written as
\begin{equation}
    \bm k = k_\parallel \bm g + k_\perp \bm{\hat z} \times  V_\text{BZ} \bm g / g^2,
\end{equation}
where we choose a rectangular Brillouin zone with $k_\parallel, k_\perp \in [-1/2, 1/2)$ and $V_\text{BZ} = 8\pi^2 / \sqrt{3} a^2$. In Fig.\ \ref{fig:hc-ws-trivial}(a) we show the WS ladder for $(m_1,m_2) = (1,0)$, i.e., the armchair direction, as a function of the field strength $\mathcal E$ and $k_\perp$ in the trivial phase. Here we only show positive quasienergies in the FZ as the spectrum is symmetric in the traceless gauge. We see that the crossings in the ladder disperse as a function of $k_\perp$ which is mainly due to the change in the center of the hybrid Wannier state. This is evident from the evolution of the Berry phases (equivalent to the Wilson loop spectrum because the bands do not cross) shown in Fig.\ \ref{fig:hc-ws-trivial}(b). Note that there is no net pumping of the Wannier center in the trivial phase as expected. We have also superimposed on the WS ladder the crossings obtained in band basis up to lowest order in interband corrections by taking into account the susceptibility. The latter is shown in Fig.\ \ref{fig:hc-ws-trivial}(c) and peaks when the integration path cuts across the zone corners where $g^{ij}/2d$ is maximal. We also show the WS ladder for a field along the zigzag direction $(m_1,m_2) = (1,-1)$ in Fig.\ \ref{fig:hc-ws-trivial}(d). In this case, the Berry phase vanishes because the field lies perpendicular to a mirror line [the $y$ axis in Fig.\ \ref{fig:honeycomb}(a)] and therefore the projected Wannier center is pinned at the origin. This is why the ladder is much less dispersive. Furthermore, the susceptibility correction is now largest at $k_\perp = 0$ since this is where both zone corners are projected on top of each other.

\begin{figure}
    \centering
    \includegraphics[width=\linewidth]{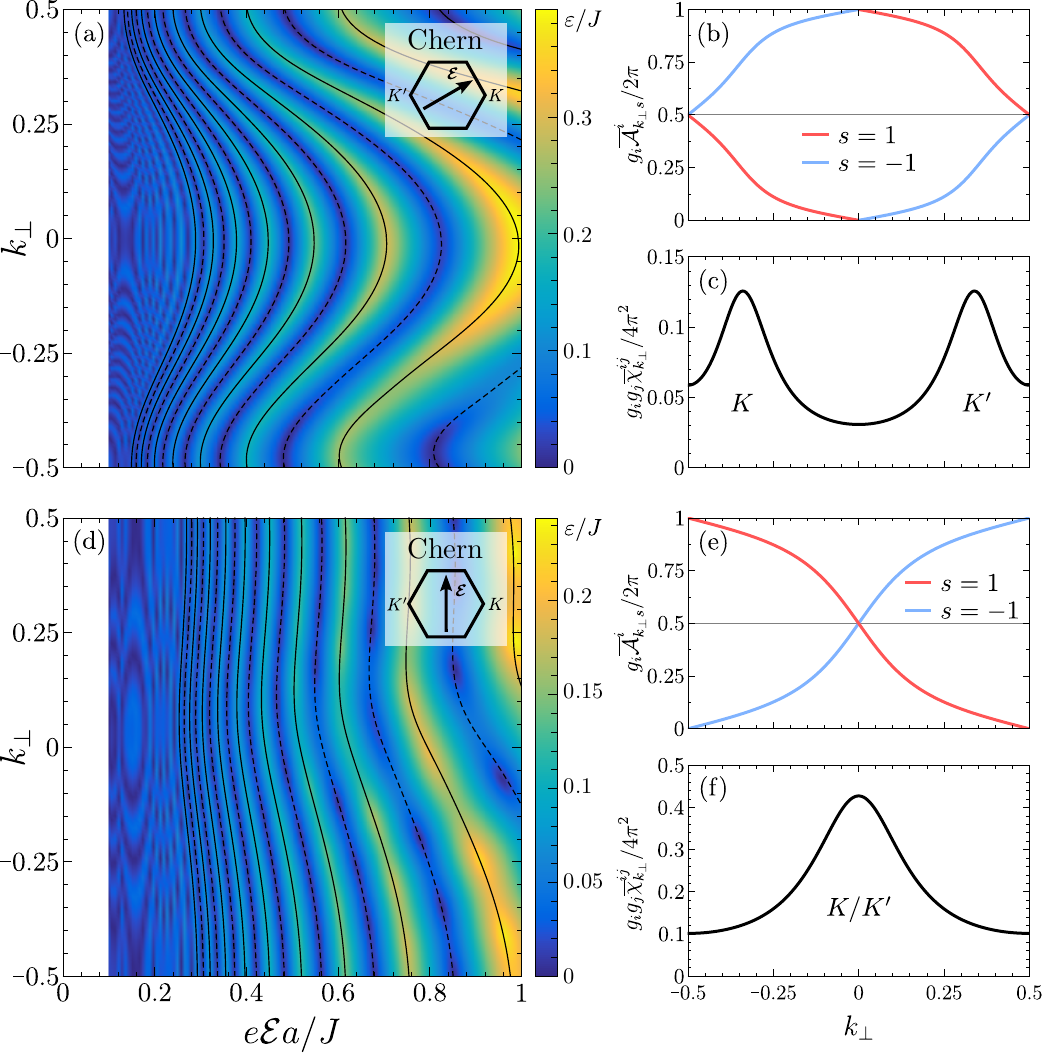}
    \caption{\textbf{WS ladder of the honeycomb lattice in the Chern phase} for $(\Delta_0,\Delta_1) = (0,0.5)$. (a) WS ladder in the FZ for $(m_1,m_2) = (1,0)$. The solid (dashed) curves give the crossings in the ladder at zero (edge of the FZ) as calculated in band basis up to 2nd order. We only show positive quasienergies since the spectrum is symmetric in traceless gauge. (b) Projected Berry phase (Wannier center) for the two bands and (c) susceptibility $\overline \chi_{k_\perp s} = s\overline \chi_{k_\perp}$ as a function of $k_\perp$ for the field in (a). (d) WS ladder, (e) Berry phase, and (f) susceptibility for $(m_1,m_2) = (1,-1)$.}
    \label{fig:hc-ws-chern}
\end{figure}

The WS ladder is qualitatively different when the Chern number $\mathcal C$ is finite, since the Berry phase winding as $k_\perp$ advances one unit is by definition given by $\mathcal C$ \cite{king-smith_theory_1993,lee_direct_2015}. In our case, we have $\mathcal C = \pm1$ such that during one cycle the Wannier center moves one cell over: $n \mapsto n \pm 1$. This can be observed in the WS ladder which is shown for $(m_1,m_2) = (1,0)$ in Fig.\ \ref{fig:hc-ws-chern}(a) and can be explicitly seen from the Berry phase winding shown in Fig.\ \ref{fig:hc-ws-chern}(b). Following the crossings in the WS ladder from $k_\perp = -1/2$ to $k_\perp = 1/2$ we end up at a different field value as the one we started from due to the shift of the Wannier center. Changing the sign of $\Delta_1$ reverses the pump and the corresponding shift in the WS ladder. A similar effect occurs in Weyl semimetals where the Weyl charge acts as a topological defect in the WS ladder \cite{kim_surface_2016}.

We next consider the current response. Before we proceed to the results, we first address how the current is constrained by the crystalline symmetries. Under a spatial symmetry $\mathcal S$, the current transforms as $\bm j(\bm{\mathcal E}) = \mathcal S \bm j(\mathcal S^{-1} \bm{\mathcal E})$. This motivates us to define the longitudinal and transverse current components as
\begin{equation}
    j_\parallel = \bm{\hat{\mathcal E}} \cdot \bm j, \qquad j_\perp = \bm{\hat z} \times \bm{\hat{\mathcal E}} \cdot \bm j,
\end{equation}
which, respectively, transform as a scalar and pseudoscalar under $\mathcal S$. Next, in order to make sense of the results obtained with the full quantum theory, we construct linear combinations of the currents that transform in the same way as their band-projected counterparts in the adiabatic theory (see Appendix \ref{app:semiclassical}). Note that these current components generally contain interband contributions, but in the limit $E_g^2/W \gg \Omega$ they become purely intraband. Other parts of the current that do not transform as intraband currents are then attributed solely to interband corrections. On the computational side, we consider commensurate fields $\bm{\mathcal E} = \mathcal E \bm g / g = \mathcal E (\cos \theta,\sin\theta)$ up to $|\bm g| = \sqrt{73} \, g_0$ in the range $\theta \in [-\pi/3,\pi/3]$. This is sufficient to reconstruct the entire angular dependence of the current from the lattice symmetries.

\begin{figure}
    \centering\includegraphics[width=\linewidth]{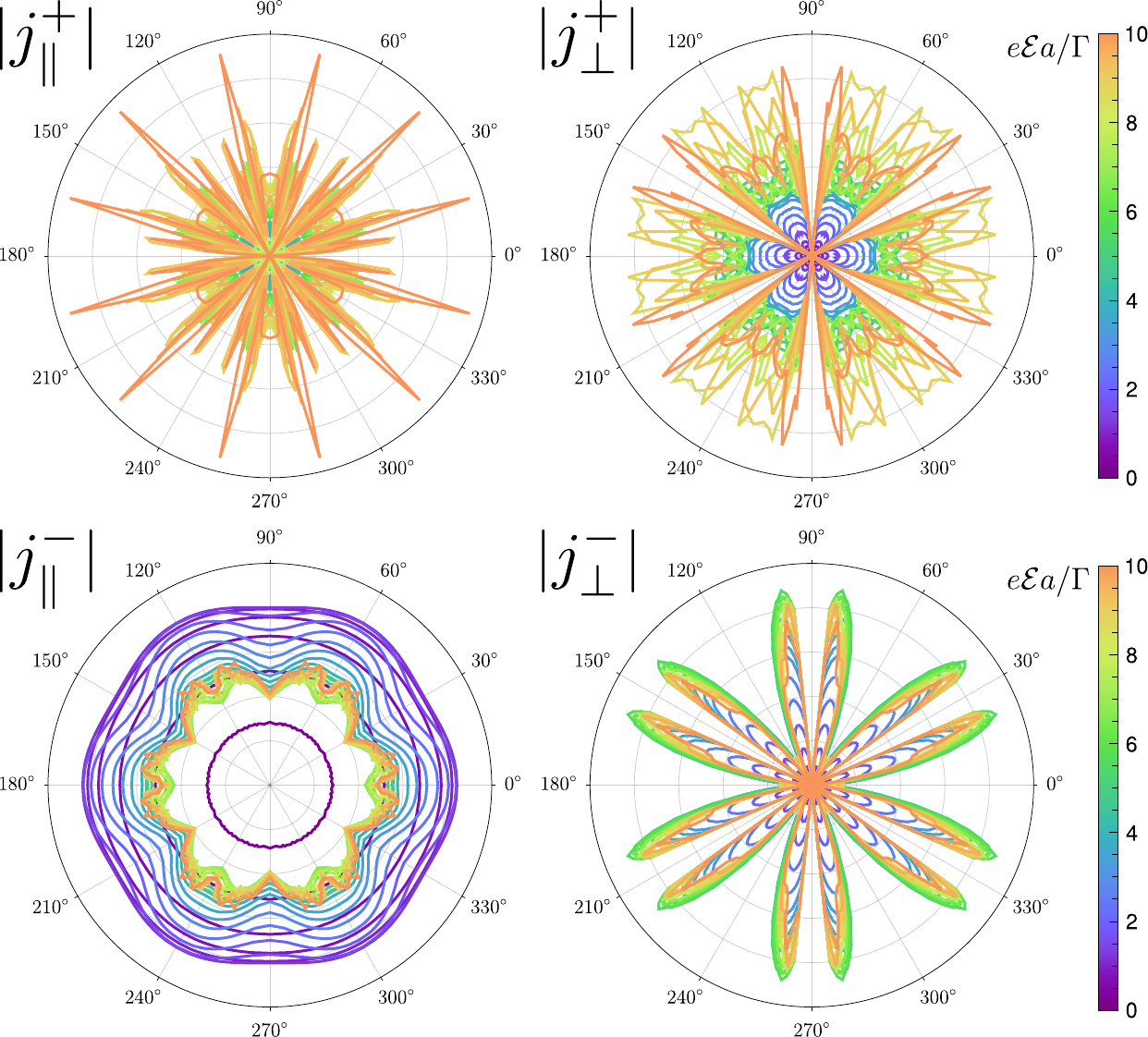}
    \caption{\textbf{Current roses for the honeycomb lattice in the trivial phase with $C_{3v}$ and $\mathcal T$ symmetry at half filling of the lower band.} Parameters are $\beta = 50$ and $\Gamma = 0.1$ with masses $(\Delta_0,\Delta_1)=(0.5,0)$ such that $\mathcal T$ is preserved. Even and odd components for $\mathcal S = 1$ are shown. Starting from $j_\parallel^+$ and going clockwise, radial ticks are spaced by $0.002$, $0.004$, $0.025$, and $0.01$ in units of $e J / \hbar a$. The odd components are nonzero because $\Delta_0$ breaks $\mathcal C_{2z}$ symmetry. The maximum number of Floquet sidebands is $N_F = 1122$. Note that the current is not fully converged for the smallest field shown.}
    \label{fig:hc_rose_trivial}
\end{figure}

In particular, we consider magnetic symmetries $\mathcal S \mathcal T$ and define the even and odd components with respect to this symmetry as follows:
\begin{equation}
    \bm j^{\pm}(\bm{\mathcal E}) \equiv \frac{\bm j(\bm{\mathcal E}) \pm \mathcal S \bm j(-\mathcal S^{-1} \bm{\mathcal E})}{2},
\end{equation}
such that $\bm j^+$ transforms as the intraband geometric current due to the Berry curvature, while $\bm j^-$ transforms as the Bloch current originating from the dispersion which is a Drude-type contribution (see Appendix \ref{app:semiclassical}). Hence, the longitudinal part of $\bm j^+$ is purely interband. For example, when $\mathcal C_{2z}$ is broken but $\mathcal T$ is conserved ($\Delta_1 = 0$) we choose $\mathcal S = 1$. We illustrate this case with the current roses \cite{de_beule_roses_2023} in Fig.\ \ref{fig:hc_rose_trivial} for $\Delta_0 = 0.5$ and $\Delta_1 = 0$.  The nodes in the transverse response are due to the presence of the mirror symmetry $\mathcal M_x$ which forbids a transverse response when the field lies along a mirror line, given here by the $y$ axis and its $\mathcal C_{3z}$ partners.

\begin{figure}
    \centering\includegraphics[width=\linewidth]{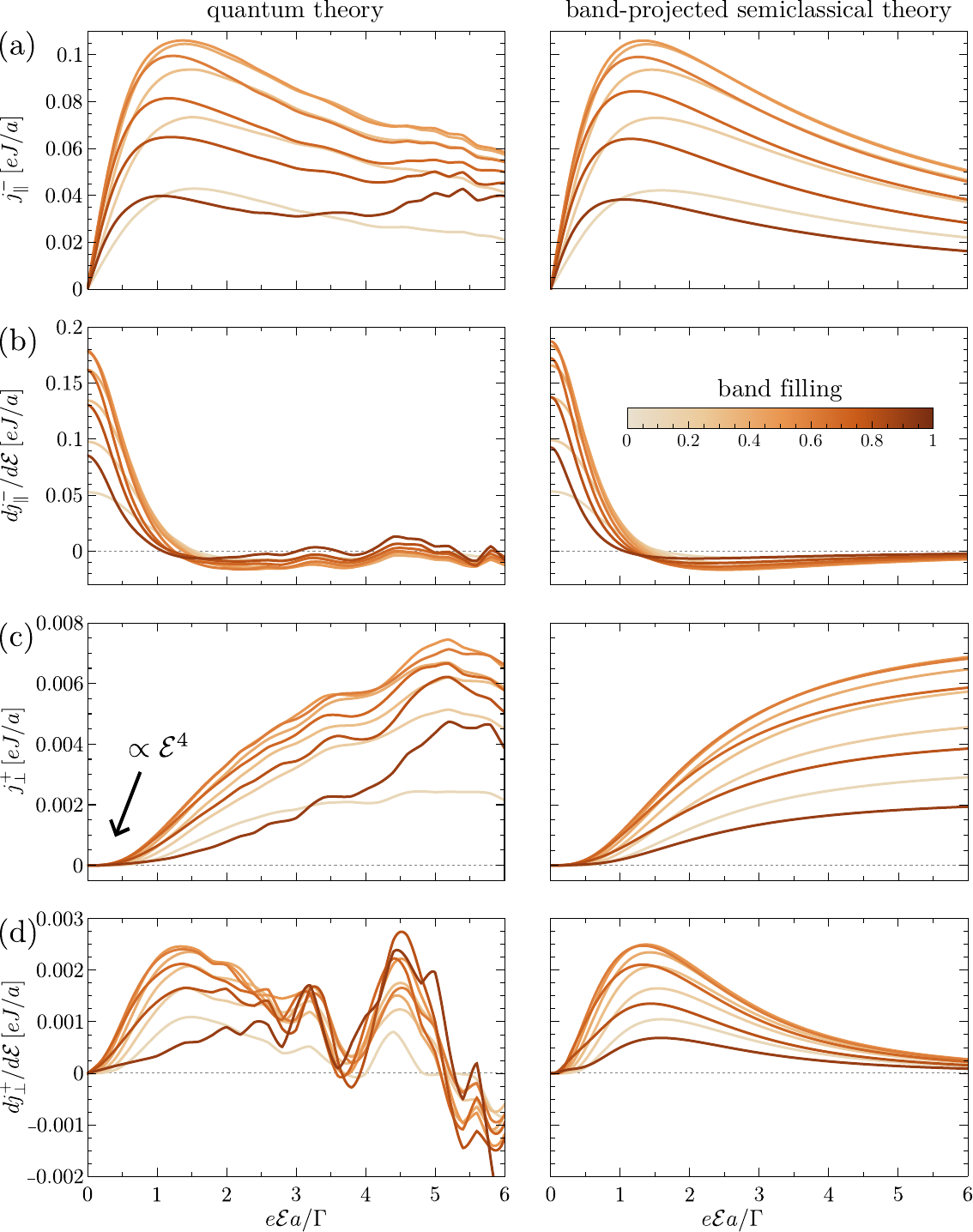}
    \caption{\textbf{Currents versus filling for the honeycomb lattice in the trivial phase with $C_{3v}$ and $\mathcal T$.} The left and right column show results for the quantum and band-projected semiclassical theory, respectively. Parameters are $\beta = 50$ and $\Gamma = 0.1$, for $\bm{\mathcal E} = \mathcal E \bm{\hat x}$ ($\theta = 0^\circ$) with masses $(\Delta_0,\Delta_1)=(0.5,0)$ such that $\mathcal T$ is preserved. (a) Current component $j_\parallel^-$ for $\mathcal S = 1$ and (b) the corresponding differential conductance. Other components vanish for $\theta = 0^\circ$. (c, d) Same for the transverse even part $j_\perp^+$. The color gives the filling of the lower band ($\nu = 0.1,0.2,\ldots,0.9$) and $N_F = 562$.}
    \label{fig:hc_filling_trivial}
\end{figure}

We also observe that the odd current components, which are mainly due to intraband Drude-like contributions, display a maximum \cite{tsu_superlattice_2005}. Indeed, at low fields $j_\parallel^-$ is isotropic and increases linearly (Ohm's law). With increasing field strength, the current becomes more anisotropic as the entire energy band is probed by the accelerated electrons, and attains an extremum for $e \mathcal E a / \Gamma \sim 1$. In this regime, complete Bloch orbits occur before a scattering event and the current decays as $1/\mathcal E$ until interband transitions become significant. The latter give rise to many oscillations from Zener resonances whose details are captured by the WS ladder. However, as long as interband contributions are small near $e \mathcal E a / \Gamma \sim 1$, the longitudinal differential conductance $dj_\parallel/d\mathcal E$ becomes negative over an extended range \cite{esaki_superlattice_1970}. This is illustrated in Fig.\ \ref{fig:hc_filling_trivial}, where we show both the current and the differential conductance for the field direction $\theta = 0^\circ$ [$(m_1,m_2)=(1,-1)$] for several fillings of the lower band. Generally, we observe that interband contributions become larger as the filling increases. This makes sense because electrons in the lower band that have a higher Fermi energy are more likely to reach the band edge, where interband coupling is stronger, before intraband relaxation can occur. Note also that the current calculated with the quantum theory is not fully converged for very small fields. This is most clearly seen in Fig.\ \ref{fig:hc_filling_trivial}(b) by comparison to the semiclassical result which becomes exact in the limit $\mathcal E \rightarrow 0$.

Moreover, the transverse differential conductance, shown in Fig.\ \ref{fig:hc_filling_trivial}(d), which is purely geometric for $\theta = 0^\circ$ when interband coupling is negligible, is cubic at low fields due to a Berry curvature hexapole in the presence of $\mathcal C_{3z}$ but broken $\mathcal C_{2z}$ \cite{zhang_higher-order_2023,de_beule_roses_2023} and shows a peak for $e \mathcal E a / \Gamma \sim 1$ indicative of the incipient plateau in the geometric current. However, unlike in the band-projected theory, the plateau itself is never reached for the chosen parameters because of interband coupling. Hence, the extremum in the transverse differential conductance at the onset of full Bloch orbits provides a more robust signature of the nominal plateau due to geometric oscillations in the band-projected theory.

\begin{figure}
    \centering\includegraphics[width=\linewidth]{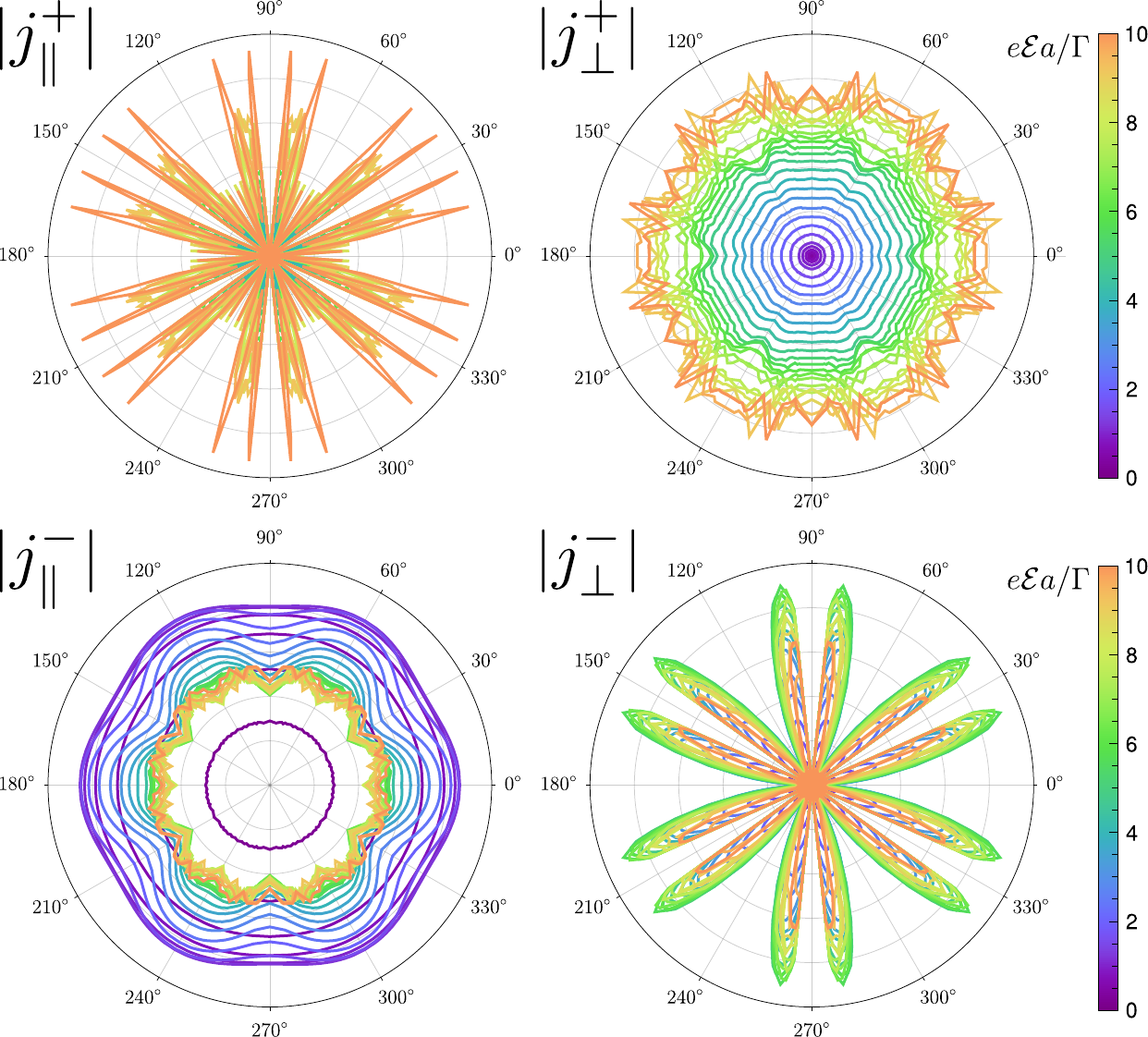}
    \caption{\textbf{Current roses for the honeycomb lattice in the Chern phase with $6 \underline{m} \underline{m}$ symmetry at half filling of the lower band.} Parameters are $\beta = 50$, $\Gamma = 0.1$ with $(\Delta_0,\Delta_1)=(0,0.5)$ such that $\mathcal C_{2z}$ is preserved. Even and odd components for $\mathcal S = \mathcal M_x$ are shown. Starting from $j_\parallel^+$ and going clockwise, radial ticks are spaced by $0.0004$, $0.014$, $0.025$, and $0.01$ in units of $e J / \hbar a$. The maximum number of Floquet sidebands is $N_F = 1122$. Note the current is not fully converged for the smallest field shown.}
    \label{fig:hc_rose_chern}
\end{figure}

On the other hand, when $\mathcal C_{2z}$ is conserved, only even current components with respect to $\mathcal S = 1$ are nonvanishing such that the above distinction becomes superfluous. Instead, we decompose the current with $\mathcal S = \mathcal M_x$ since $\mathcal M_x \mathcal T$ is always conserved regardless of the sublattice or Haldane masses. This is shown in Fig.\ \ref{fig:hc_rose_chern} for the case with $6\underline{m}\underline{m}$ symmetry in the Chern phase where we take $\Delta_0 = 0$ and $\Delta_1 = 0.5$. While the pure interband component $j_\parallel^+$ and the odd components with respect to $\mathcal M_x \mathcal T$ are qualitatively the same from the components defined with respect to $\mathcal T$ for the case with $C_{3v}$ and $\mathcal T$ symmetry (see Fig.\ \ref{fig:hc_rose_trivial}), the transverse even component $j_\perp^+$ looks very different. This is because of the finite Chern number which gives rise to a dominant linear contribution to the current that is isotropic and given by $\left( e^2 \mathcal E / V_c \hbar \right) \sum_{\bm r} f_{\bm r}^0 \Omega_{-\bm r} = \left( e^2 V_c \mathcal E / 2\pi h \right) \int_\text{BZ} d^2 \bm k f_{\bm k}^0 \Omega_{\bm k}$ which is the Hall conductance. We note that in Fig.\ \ref{fig:hc_rose_chern}, the seemingly nonlinear bunching at small fields for $j_\perp^+$ is due to our choice for the $\mathcal E$ grid which has twice the number of points for $| e \mathcal E a / \Gamma | \leq 1.2$.

\begin{figure}
    \centering\includegraphics[width=\linewidth]{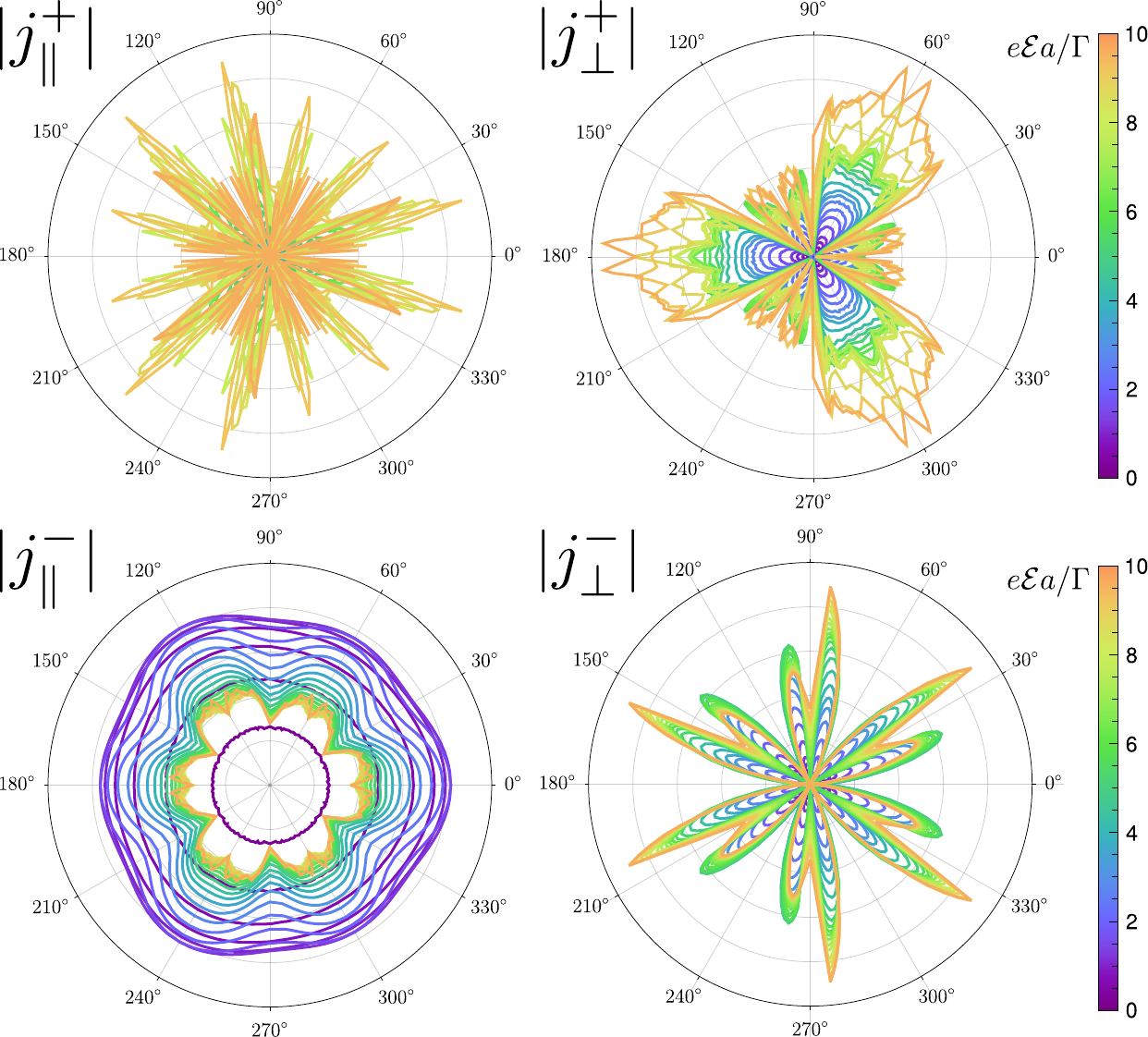}
    \caption{\textbf{Current roses for the honeycomb lattice in the trivial phase with $3 \underline{m}$ symmetry at half filling of the lower band.} Parameters are $\beta = 50$, $\Gamma = 0.1$ with $(\Delta_0,\Delta_1)=(0.75,0.25)$ such that $\mathcal T$ is broken. Even and odd components for $\mathcal S = \mathcal M_x$ are shown. Starting from $j_\parallel^+$ and going clockwise, radial ticks are spaced by $0.0011$, $0.003$, $0.025$, and $0.014$ in units of $e J / \hbar a$. The maximum number of Floquet sidebands is $N_F = 1002$. Note that the current is not fully converged for the smallest field shown.}
    \label{fig:hc_rose_trivial_brokenT}
\end{figure}

Finally, we consider the case with $3\underline{m}$ symmetry for which both $\mathcal C_{2z}$ and $\mathcal T$ are broken. Here we consider the trivial phase with $\Delta_0 = 0.75$ and $\Delta_1 = 0.25$. Similar to before, since $\mathcal M_x \mathcal T$ remains conserved, we decompose the current in terms of even and odd components with respect to this symmetry. Since both $\mathcal T$ and $\mathcal C_{6z}$ are broken, the current roses shown in Fig.\ \ref{fig:hc_rose_trivial_brokenT} only retain $\mathcal C_{3z}$ symmetry. Moreover, the even transverse component that is due to the Berry curvature in the intraband limit now contains both contributions from broken $\mathcal C_{2z}$ ($\Delta_0$) and broken $\mathcal T$ ($\Delta_1$) which have a different angular dependence such that they either add or substract giving rise to the $j_\perp^+$ rose shown in Fig.\ \ref{fig:hc_rose_trivial_brokenT}. We further note that because $\mathcal C_{2z}$ and $\mathcal T$ are both broken in this case, interband contributions are generally more important. This is because the lowest-order interband contribution to the current, i.e., due to the electric susceptibility, vanishes in the presence of either of these symmetries \cite{gao_field_2014}. It would be interesting to study the current nonperturbatively for a system with broken $\mathcal C_{2z}$ (or $\mathcal P = \mathcal M_z \mathcal C_{2z}$) and $\mathcal T$ symmetry but still conserves their combination $\mathcal C_{2z} \mathcal T$ such that the Berry curvature vanishes. In this case, the even current component $\bm j^-$ (not just the longitudinal part) is entirely due to interband coupling. Moreover, at order $\mathcal E^2$, this response yields a measure for the quantum metric of the occupied band \cite{gao_field_2014,wang_quantum-metric-induced_2023,gao_quantum_2023}. However, this would require a lattice model that is either strongly anisotropic or has more than two bands and therefore we leave this for future work.
       
\section{Conclusions} \label{sec:conclusions}

In this work, we considered the full quantum theory of the electric current response to a static uniform electric field of noninteracting fermions on a lattice in the regime of Bloch oscillations ($\omega_B \tau \gg 1$). As such, our theory is nonperturbative in the interband coupling. We achieve this by first mapping the physical \emph{time-independent} problem in $D$ spatial dimensions to a \emph{time-dependent} problem in $D-1$ spatial dimensions by treating the field in temporal gauge. In this gauge, the longitudinal momentum becomes a gauge degree of freedom which can be absorbed in the time origin. Moreover, when the electric field is commensurate to the lattice, i.e., when it lies along a reciprocal lattice vector (equivalently perpendicular to a lattice plane), Bloch oscillations in the semiclassical theory become periodic and likewise the Bloch Hamiltonian in the quantum theory becomes time periodic up to a unitary transformation. The latter is a consequence of working in periodic gauge for the total Bloch wave function and properly takes into account the sublattice positions. By using a modified Floquet \emph{ansatz} one obtains a problem in $D-1$ spatial dimensions with an additional synthetic Floquet dimension which can be interpreted as the spatial direction that lies longitudinal to the commensurate electric field. The corresponding Floquet quasienergies then yield the familiar Wannier-Stark ladder beyond the single-band approximation.

To obtain the current, we coupled the system to a reservoir using the Floquet-Keldysh Green's function formalism. For simplicity, we considered a reservoir that conserves the periodicity of the lattice in the wide-band limit. In this case, the self-energy due to the bath is diagonal in momentum space and Floquet space yielding a relatively simple expression for the current. Moreover, in order to study the onset of interband contributions to the current, we also derived expressions for the band-resolved currents using the Floquet representation of band projectors. We then used this formalism to study the current response in $D = 1$ and $D = 2$ spatial dimensions.

In 1D for a single-band model, we find nearly perfect agreement with the band-projected semiclassical theory apart from small oscillations related to the nature of the bath and which vanish in the limit $\Gamma / W \rightarrow 0$, where $W$ is the bandwidth and $\Gamma$ is the system-bath coupling. For a two-band model in 1D, deviations from the semiclassical theory arise due to interband currents (Zener transitions) which become significant for $e \mathcal E a \sim E_g^2 / W$. Moreover, as we increase the chemical potential and the lower band becomes more filled, one finds that interband contributions become more important since Zener transitions are more likely to occur before intraband relaxation. Hence, in the regime of full Bloch orbits, the validity of the band-projected theory is estimated as: $a^2 E_g / | \mathcal A_\text{inter} |^2 \gg \Omega \gg \Gamma$ where $\mathcal A_\text{inter}$ is the interband Berry connection.

In 2D, we considered a honeycomb lattice with both sublattice ($\mathcal C_{2z}$ breaking) and Haldane mass ($\mathcal T$ breaking) terms that exhibits both trivial gapped and Chern phases. In the latter case, the finite Chern number can be expressed as the winding number of the Berry phase along the momentum direction transverse to the field direction, giving rise to a net pumping of the Wannier center across the real-space cell. This can be observed directly in the Wannier-Stark ladder which encodes the Berry phase winding through the polarization as one traverses the zone. We proceeded to calculate the current response for a honeycomb model. In order to compare our results to the semiclassical theory, we decomposed the current into parts that transform in the same way as their semiclassical counterparts under a magnetic symmetry $\mathcal S \mathcal T$. Most interestingly, when $\mathcal T$ is conserved but $\mathcal C_{2z}$ is broken, we find that unless the band gap opened by the sublattice mass is very large, the transverse geometric response does not plateau as a function of the field strength as was predicted in the band-projected theory. Instead, interband contributions become significant before the plateau is reached. However, we find that the corresponding peak in the differential conductance due to the incipient plateau provides a more robust probe of geometric oscillations.   

\begin{acknowledgements}
We acknowledge interesting and fruitful discussions with V.~T.\ Phong, M.\ Claassen, and C.~L.\ Kane. C.~D.~B.\ and E.~J.~M.\ are supported by the Department of Energy under grant DE-FG02-84ER45118. S.~G.\ and S.~T.\ acknowledge funding from the NSF GRFP under grant DGE-1845298.
\end{acknowledgements}

\appendix

\section{Length gauge} \label{app:length_gauge}

Here we give a short overview of the transformation between velocity gauge and length gauge. Like velocity gauge, length gauge is an incomplete gauge that becomes fixed in the electric-dipole approximation. In length gauge, one describes a uniform electric field $\bm{\mathcal E}(t)$ with an electrostatic potential $\varphi(\bm r,t) = -\bm{\mathcal E}(t) \cdot \bm r$. The upside of length gauge is that the Hamiltonian itself remains time independent, but the downside is that it breaks translational symmetry. We can transform the Hamiltonian in velocity gauge given by Eq.\ \eqref{eq:Hsys} to length gauge with the unitary transformation $U(t) = e^{iS(t)}$ where
\begin{equation}
    S(t) = e \bm A(t) \cdot \sum_{\bm r,a} ( \bm r + \bm r_a ) c_{\bm ra}^\dag c_{\bm ra}.
\end{equation}
The Hamiltonian transforms as
\begin{equation}
    H \rightarrow \breve H = U H U^\dag + i \dot U U^\dag,
\end{equation}
where the second term on the right-hand side is straightforward to evaluate. To deal with the first term, consider
\begin{equation}
    e^{iS(t)} c_{\bm ra}^\dag c_{\bm r'b} e^{-iS(t)},
\end{equation}
which can be obtained from a special case of the Baker-Campbell-Hausdorff formula given by
\begin{equation}
    e^{iX} Y e^{-iX} = Y + i[X, Y] + \frac{i^2}{2} [X, [X, Y]] + \cdots,
\end{equation}
with $X = S(t)$ and $Y = c_{\bm ra}^\dag c_{\bm r'b}$. From
\begin{equation}
    [ c_{\bm r''c}^\dag c_{\bm r''c} , c_{\bm ra}^\dag c_{\bm r'b} ] = c_{\bm ra}^\dag c_{\bm r'b} \left( \delta_{\bm r\bm r''} \delta_{ac} - \delta_{\bm r'\bm r''} \delta_{ab} \right),
\end{equation}
we find $[X, Y] = Y e \bm A(t) \cdot \left( \bm r - \bm r' + \bm r_{ab} \right)$. We obtain
\begin{equation}
    e^{iS(t)} c_{\bm ra}^\dag c_{\bm r'b} e^{-iS(t)} = c_{\bm ra}^\dag c_{\bm r'b} e^{i e \bm A(t) \cdot \left( \bm r - \bm r' + \bm r_{ab} \right)},
\end{equation}
and the transformed Hamiltonian becomes
\begin{equation}
    \breve H = H_0 + e \bm{\mathcal E}(t) \cdot \sum_{\bm r,a} ( \bm r + \bm r_a ) c_{\bm ra}^\dag c_{\bm ra},
\end{equation}
where the last term gives the potential energy of charge $-e$ fermions on the lattice in a uniform field $\bm{\mathcal E}(t)$.

\section{Semiclassical band-projected theory of Bloch and geometric oscillations} \label{app:semiclassical}

Here we give a short review of the semiclassical band-projected transport theory of Bloch and geometric oscillations \cite{tsu_superlattice_2005,phong_quantum_2023,de_beule_roses_2023,de_beule_berry_2023}. In this section, we restore $\hbar$ for consistency with existing literature. For a crystal in a uniform electric field, the Boltzmann transport equation for the occupation function in the relaxation-time approximation has the steady-state solution \cite{ashcroft_solid_1976}:
\begin{equation}
    \begin{aligned}
        & f(t) = f^0(t) \\
        & + \frac{e}{\hbar} \int_{-\infty}^t dt' \, \exp \left( - \int_{t'}^t \frac{ds}{\tau(s)} \right) \nabla_{\bm k} f^0(t') \cdot \bm{\mathcal E}(t'),
    \end{aligned}
\end{equation}
with $\tau(t) = \tau[\bm k(t)]$ the momentum-relaxation time. For a static field and a constant relaxation time, this can be solved exactly \cite{lebwohl_electrical_2003,de_beule_roses_2023},
\begin{equation} \label{eq:f}
    f(t) = f_{\bm k(t)} = \sum_{\bm r} \frac{f_{\bm r}^0 \, e^{i \bm k(t) \cdot \bm r}}{1 - i e \tau \bm r \cdot \bm{\mathcal E} / \hbar},
\end{equation}
where the sum runs over lattice vectors $\bm r$ and $f^0_{\bm r} = [ V_c / \left( 2 \pi \right)^d ] \int_\text{BZ} d^d\bm k \, f_{\bm k}^0 \, e^{-i \bm k \cdot \bm r}$ are lattice Fourier components of the Fermi-Dirac distribution. The current in the band-projected theory is given by $\bm j_\text{Bloch} + \bm j_\text{geom}$ with
\begin{align}
    \bm j_\text{Bloch} & = -\frac{e}{\hbar} \int_{\bm k} \, f_{\bm k} \nabla_{\bm k} \varepsilon_{\bm k}, \\
    \bm j_\text{geom} & = -\bm{\mathcal E} \times \frac{e^2}{\hbar} \int_{\bm k} \, f_{\bm k} \bm{\Omega}_{\bm k},
\end{align}
where $\int_{\bm k} \equiv \int_\text{BZ} d^D{\bm k} / (2\pi)^D$, $\varepsilon_{\bm k}$ is the energy band, and $\bm \Omega_{\bm k} = \nabla_{\bm k} \times \bm{\mathcal A}_{\bm k}$ is the Berry curvature in periodic gauge written as a pseudovector. Plugging in the occupation function from Eq.\ \eqref{eq:f} yields
\begin{align}
    \bm j_\text{Bloch} & = \frac{ie}{V_c \hbar} \sum_{\bm r} \frac{\bm r f_{\bm r}^0 \varepsilon_{-\bm r}}{1 - i e \tau \bm r \cdot \bm{\mathcal E} / \hbar}, \\
    \bm j_\text{geom} & = - \bm{\mathcal E} \times \frac{e^2}{V_c \hbar} \sum_{\bm r} \frac{f_{\bm r}^0 \bm{\Omega}_{-\bm r}}{1 - i e \tau \bm r \cdot \bm{\mathcal E} / \hbar},
\end{align}
with $V_c$ the unit cell volume, and where $\varepsilon_{\bm r}$ and $\bm{\Omega}_{\bm r}$ are lattice Fourier components of the energy band and Berry curvature, respectively. In particular, note that $f^0_{\bm 0}$ gives the filling fraction of the band and for $D=2$ we have $\Omega_{\bm 0} = V_c \mathcal C / 2 \pi$ with $\mathcal C$ the Chern number.

It is instructive to rewrite the Bloch current as
\begin{equation}
    \bm j_\text{Bloch} = \frac{ie}{V_c \hbar} \left( \sum_{\bm r \cdot \bm{\mathcal E} = 0} \bm r f_{\bm r}^0 \varepsilon_{-\bm r} + \sum_{\bm r \cdot \bm{\mathcal E} \neq 0} \frac{\bm r f_{\bm r}^0 \varepsilon_{-\bm r}}{1 - i e \tau \bm r \cdot \bm{\mathcal E} / \hbar} \right),
\end{equation}
where the first term vanishes in the presence of time-reversal or spatial inversion symmetry. Each of these symmetries individually imply that $\varepsilon_{\bm r} = \varepsilon_{-\bm r}$ and $f_{\bm r}^0 = f_{-\bm r}^0$ are real, such that $\bm j_\text{Bloch}(-\bm{\mathcal E}) = -\bm j_\text{Bloch}(\bm{\mathcal E})$. Restricting to $D=2$, we define $\bm j_\text{geom}  = j_\text{geom} \bm{\hat z} \times \bm{\hat{\mathcal E}}$ with \begin{equation}
    j_\text{geom} = \frac{e^2 \mathcal E}{V_c \hbar} \left( \sum_{\bm r \cdot \bm{\mathcal E} = 0} f_{\bm r}^0 \Omega_{-\bm r} + \sum_{\bm r \cdot \bm{\mathcal E} \neq 0} \frac{f_{\bm r}^0 \Omega_{-\bm r}}{1 - i e \tau \bm r \cdot \bm{\mathcal E} / \hbar} \right),
\end{equation}
where now the first term only vanishes in the presence of time-reversal  $\mathcal T$. Indeed, time-reversal symmetry implies that $\Omega_{\bm r} = -\Omega_{-\bm r}$ is imaginary, and $\bm j_\text{geom}$ is even in $\bm{\mathcal E}$ in this case. On the other hand, inversion symmetry gives real $\Omega_{\bm r} = \Omega_{-\bm r}$ and $\bm j_\text{geom}$ is odd in $\bm{\mathcal E}$ in this case.

The transformation properties of the Bloch and geometric current under a general crystalline symmetry $\mathcal S$ follow similarly from $\varepsilon_{\bm r} = \varepsilon_{\mathcal S \bm r}$, $f^0_{\bm r} = f^0_{\mathcal S \bm r}$, and $\Omega_{\bm r} = \det(\mathcal S) \Omega_{\mathcal S \bm r}$, while a magnetic symmetry $\mathcal S \mathcal T$ implies that $\varepsilon_{\bm r} = \varepsilon_{-\mathcal S \bm r}$, $f^0_{\bm r} = f^0_{-\mathcal S \bm r}$, and $\Omega_{\bm r} = - \det(\mathcal S) \Omega_{-\mathcal S \bm r}$. With these relations, one can demonstrate readily that the currents transform as 
\begin{align}
    \mathcal S: \quad & \bm j(\bm{\mathcal E}) = \mathcal S \bm j(\mathcal S^{-1} \bm{\mathcal E}), \\
    \mathcal S \mathcal T: \quad &
    \begin{aligned}
        \bm j_\text{Bloch}(\bm{\mathcal E}) & = - \mathcal S \bm j_\text{Bloch}(-\mathcal S^{-1} \bm{\mathcal E}), \\
        \bm j_\text{geom}(\bm{\mathcal E}) & = \mathcal S \bm j_\text{geom}(-\mathcal S^{-1} \bm{\mathcal E}).
    \end{aligned}
\end{align}
Moreover, defining the longitudinal and transverse components,
\begin{equation}
    j_\parallel = \bm{\hat{\mathcal E}} \cdot \bm j, \qquad j_\perp = \bm{\hat z} \times \bm{\hat{\mathcal E}} \cdot \bm j,
\end{equation}
we find that
\begin{align}
    \mathcal S: \quad &
    \begin{aligned}
        j_\parallel(\bm{\mathcal E}) & = j_\parallel(\mathcal S^{-1} \bm{\mathcal E}), \\
        j_\perp(\bm{\mathcal E}) & = \det(\mathcal S) j_\perp(\mathcal S^{-1} \bm{\mathcal E}),
    \end{aligned} \\[1.5mm]
    \mathcal S \mathcal T: \quad &
    \begin{aligned}
        j_\text{Bloch}^\parallel(\bm{\mathcal E}) & = j_\text{Bloch}^\parallel(-\mathcal S^{-1} \bm{\mathcal E}), \\
        j_\text{Bloch}^\perp(\bm{\mathcal E}) & = \det(\mathcal S) j_\text{Bloch}^\perp(-\mathcal S^{-1} \bm{\mathcal E}), \\
        j_\text{geom}(\bm{\mathcal E}) & = -\det(\mathcal S) j_\text{geom}(-\mathcal S^{-1} \bm{\mathcal E}).
    \end{aligned}
\end{align}

As an example, consider a Bloch band in $D=1$ with time-reversal or inversion symmetry. Now we only have a longitudinal response
\begin{align}
    j_\text{Bloch} & = \frac{ie}{\hbar} \sum_{n=1}^\infty n f^0_n \varepsilon_n \left( \frac{1}{1 - in\Omega \tau} - \frac{1}{1 + in\Omega \tau} \right) \\
    & = -\frac{2e \Omega \tau}{\hbar} \sum_{n=1}^\infty \frac{ n^2 f^0_n \varepsilon_n}{1 + \left( n\Omega \tau \right)^2},
\end{align}
with $\Omega = eEa/\hbar$ where $a$ is the lattice constant. Specifically, for a linear chain with nearest-neighbor hopping amplitude $J$, we find
\begin{equation} \label{eq:j_chain_sc}
    j_\text{Bloch} = -\frac{2eJf_a^0}{\hbar} \frac{\Omega \tau}{1 + ( \Omega \tau )^2},
\end{equation}
which is shown in Fig.\ \ref{fig:j_1band}. Here the lattice Fourier transform of the Fermi-Dirac distribution is given by [substituting $u = \cos(ka)$]
\begin{align}
    f_a^0 & = \frac{a}{2\pi} \int_{-\pi/a}^{\pi/a} dk \, f^0(\varepsilon_k) \cos(ka) \\
    & = \frac{1}{\pi} \int_{-1}^1 du\, \frac{u f^0(2Ju)}{\sqrt{1-u^2}} \\
    & \simeq -\frac{\sgn(J)}{\pi} \, \sqrt{1 - \left( \frac{\mu}{2 J} \right)^2} \, \theta \left( 1 - \left| \frac{\mu}{2J} \right| \right),
\end{align}
where the last line holds for $\beta|J| \gg 1$ and whose magnitude is largest at half filling ($\mu = 0$).

\section{Band basis} \label{app:bandbasis}

As an alternative to the orbital basis in velocity gauge, we can instead work in the instantaneous band basis by writing \cite{krieger_time_1986,phong_quantum_2023}
\begin{equation}
    \left| \Phi_{\bm k}(t) \right> = \sum_{s=1}^q a_{\bm ks}(t) \left| u_{\bm ks}(t) \right>,
\end{equation}
with $\left| u_{\bm ks}(t) \right> = \left| u_s[\bm k + e \bm A(t)] \right>$ such that
\begin{equation}
    \mathcal H(\bm k,t) \left| u_{\bm ks}(t) \right> = E_{\bm ks}(t) \left| u_{\bm ks}(t) \right>,
\end{equation}
with $E_{\bm ks}(t)$ the instantaneous band energy and normalization $\langle u_{\bm ks}(t) | u_{\bm ks'}(t) \rangle = \delta_{ss'}$. The dynamics is determined by Eq.\ \eqref{eq:schrodinger}, which yields
\begin{equation} \label{eq:band}
    i \partial_t a_{\bm k}(t) = \mathscr H(\bm k,t) a_{\bm k}(t),
\end{equation}
with $a_{\bm k} = \left( a_{\bm k1} , \ldots , a_{\bm kq} \right)^t$ and
\begin{equation}
    \mathscr H^{ss'}(\bm k,t) = \delta_{ss'} E_{\bm ks}(t) + e \bm{\mathcal E}(t) \cdot \bm{\mathcal A}_{ss'}(\bm k,t),
\end{equation}
is the Hamiltonian in the instantaneous band basis. Here we used
\begin{align}
    \hspace{-1mm}
    \left< u_{\bm ks}(t) \right| i\partial_t \left| u_{\bm ks'}(t) \right> & = (e\partial_t \bm A) \cdot \left< u_{\bm ks}(t) \right| i\partial_{\bm k} \left| u_{\bm ks'}(t) \right> \\
    & = -e\bm{\mathcal E}(t) \cdot \bm{\mathcal A}_{ss'}(\bm k,t),
\end{align}
where $\bm{\mathcal A}_{ss'}(\bm k,t) = \bm{\mathcal A}_{ss'}[\bm k + e\bm A(t)]$ is the instantaneous Berry connection. To find an approximate solution, we first rewrite Eq.\ \eqref{eq:band} by an instantaneous diagonalization, defined by $a_{\bm k}(t) = \mathscr U(\bm k,t) \breve a_{\bm k}(t)$ which yields
\begin{equation}
    i \partial_t \breve a_{\bm k}(t) = \left[ \mathscr D(\bm k,t) - i \mathscr U^\dag \dot{\mathscr U} \right] \breve a_{\bm k}(t), \label{eq:wow}
\end{equation}
where $\mathscr D(\bm k,t) = \mathscr U^\dag(\bm k,t) \mathscr H(\bm k,t) \mathscr U(\bm k,t)$ is a diagonal matrix and
\begin{equation}
    -i \mathscr U^\dag \dot{\mathscr U} = ie \bm{\mathcal E}(t) \cdot \mathscr U^\dag \nabla_{\bm k} \mathscr U - i \dot{\bm{\mathcal E}} \cdot \mathscr U^\dag \nabla_{\bm{\mathcal E}} \mathscr U,
\end{equation}
is an effective connection for the Hamiltonian in band basis which is at least third order in $\mathcal E$ and $\dot{\mathcal E}$. Up to second order, we can thus approximate the right-hand side of Eq.\ \eqref{eq:wow} by standard nondegenerate perturbation theory. This yields instantaneous eigenvalues,
\begin{equation}
    \begin{aligned}
        \lambda_{\bm ks}(t) & = E_{\bm ks}(t) + e \bm{\mathcal E}(t) \cdot \bm{\mathcal A}_{\bm ks}(t) \\
        & + \sum_{s' \neq s} \frac{[ e \bm{\mathcal E}(t) \cdot \bm{\mathcal A}_{ss'}(\bm k,t) ] [ e \bm{\mathcal E}(t) \cdot \bm{\mathcal A}_{s's}(\bm k,t) ]}{E_{\bm ks}(t) - E_{\bm ks'}(t)},
    \end{aligned}
\end{equation}
with $\bm{\mathcal A}_{\bm ks}(t) = \bm{\mathcal A}_{ss}(\bm k,t)$ the instantaneous intraband Berry connection. Hence we obtain
\begin{equation}
    \breve a_{\bm ks}(t) \approx e^{-i \int_{0}^t dt' \, \lambda_{\bm ks}(t')} \breve a_{\bm ks}(0),
\end{equation}
which is valid up to second order. Here the coefficients $\breve a_{\bm ks}(t)$ are superpositions of the original $a_{\bm ks}(t)$ such that $\mathscr H(\bm k,t)$ becomes diagonal. Moreover, for a commensurate static electric field $\bm{\mathcal E} = \mathcal E \bm g / g$, the quasienergies are defined by $\breve a_{\bm ks}(t+T) = \exp\left( -i\varepsilon_{\bm ks} T \right) \breve a_{\bm ks}(t)$ which yields
\begin{widetext}
\begin{align} \label{eq:wow2}
    \varepsilon_{\bm k_\perp s,n} & = \frac{1}{T} \int_0^T dt \, E_{\bm ks}(t) + \Omega \left[ n + \frac{g_i}{2\pi} \frac{1}{T} \int_0^T dt \left( \mathcal A_{\bm ks}^i(t) + \Omega \, \frac{g_j \chi_{\bm k s}^{ij}(t)}{2\pi} \right) \right] \\
    & = \overline{E}_{\bm k_\perp s} + \Omega \left[ n + \frac{g_i}{2\pi} \left( \overline{\mathcal A}^i_{\bm k_\perp s} + \Omega \, \frac{g_j \overline \chi^{ij}_{\bm k_\perp s}}{2\pi} \right) \right],
\end{align}
\end{widetext}
where summation over repeated indices is implied, $\Omega = 2\pi eE/g$, and
\begin{equation} \label{eq:chi2}
    \chi_{\bm ks}^{ij}(t) = \sum_{s' \neq s} \frac{\mathcal A_{ss'}^i(\bm k,t) \mathcal A_{s's}^j(\bm k,t)}{E_{\bm ks}(t) - E_{\bm ks'}(t)}, 
\end{equation}
where $\overline \chi^{ij}_{\bm k_\perp s}$ is the state-resolved (static) electric susceptibility \cite{komissarov_quantum_2023}. We see that the lowest-order interband correction is quadratic in the electric field and corresponds to the electric susceptibility which gives a field-induced shift of the polarization \cite{gao_field_2014,kane_zener_1960}. Energy gaps in the WS ladder due to interband Zener transitions thus only arise at higher orders in the field, which is corroborated by our numerical results in the main text.

As a minimal example, we consider a two-band model with Bloch Hamiltonian $\mathcal H(\bm k) = d_0(\bm k) \sigma_0 + \bm d(\bm k) \cdot \bm \sigma$ and bands $E_{\bm ks} = d_0(\bm k) + s d(\bm k)$ ($s=\pm1$) where $d = |\bm d|$. In this case, Eq.\ \eqref{eq:chi2} simplifies to
\begin{equation}
    \chi_{\bm ks}^{ij} = \frac{g^{ij}_{\bm k}}{2sd(\bm k)},
\end{equation}
with $g^{ij}_{\bm k} = \left( \partial_{k_i} \bm n \right) \cdot \left( \partial_{k_j} \bm n \right) / 4$ the quantum metric of either band and where $\bm n = \bm d / d$.

\section{Keldysh Formalism} \label{app:keldysh}

Here we review the essential parts of the Keldysh formalism for the treatment of responses. In particular, we present a derivation for the expression $G^< = G^R \Sigma^< G^A$, which gives the lesser Green's function in terms of dressed Green's functions and the lesser self energy, and we present an expression for $\Sigma^<$ in terms of the advanced, retarded, and Keldysh self energies.

\subsection{Lesser Green's Function}

We begin with the Dyson equation
\begin{equation}
    G = G_0 + G_0 \Sigma G,
\end{equation}
where $\Sigma$ is the irreducible self energy, which is composed of one-part irreducible diagrams \cite{coleman_introduction_2015}. To proceed, we consider $P = A B C$ for any operators $A$, $B$, $C$ on the (Keldysh) contour. Then the corresponding Langreth rule that takes operators on the contour to operators in real time is \cite{jauho_introduction_2006}
\begin{equation}
    P^< = A^R B^R C^< + A^R B^< C^A + A^< B^A C^A.
\end{equation}
Now for the problem of interest $P = G_0 \Sigma G$, giving
\begin{equation}
    G^< = G_0^< + G_0^R\Sigma^R G^< + G_0^R \Sigma^< G^A + G_0^<\Sigma^A G^A,
\end{equation}
where the first term is the boundary term at the initial time. This expression still contains bare Green's functions. In order to eliminate these, we solve iteratively for the dressed lesser Green's function $G^<$. This yields an expression in terms of a term linear in bare Green's functions, a term with only dressed Green's functions, and a term with infinite-order bare Green's functions
\begin{align}
    G^< & = (1+G^R\Sigma^R)G_0^<(1+\Sigma^A G^A)\nonumber\\
    & \quad + G^R \Sigma^< G^A + (G_0^R \Sigma^R)^\infty G^<.
\end{align}
In the steady state, the first term vanishes \cite{jauho_introduction_2006} and for $\Gamma$ small compared to the bandwidth such that $|\!\det(G_0^R \Sigma^R)|<1$, the last term also vanishes and so
\begin{equation}
    G^< = G^R \Sigma^< G^A.
\end{equation}

\subsection{Self energies}

In the Keldysh formalism, the self energies have the same structure as the Green's functions \cite{rammer_quantum_1986}
\begin{align}
    \Sigma^R(t_1,t_2) & = \Theta(t_1-t_2) \left[ \Sigma^>(t_1,t_2) - \Sigma^<(t_1,t_2) \right], \\
    \Sigma^A(t_1,t_2) & = -\Theta(t_2-t_1) \left[ \Sigma^>(t_1,t_2) - \Sigma^<(t_1,t_2) \right], \\
    \Sigma^K(t_1,t_2) & = \Sigma^>(t_1,t_2) + \Sigma^<(t_1,t_2),
\end{align}
with $\Theta(t)$ the Heaviside step function. Hence we have
\begin{gather}
    \Sigma^R-\Sigma^A = \Sigma^> - \Sigma^<, \\
    \Sigma^R-\Sigma^A-\Sigma^K = -2\Sigma^<,
\end{gather}
and thus
\begin{equation}
    \Sigma^< = \frac{\Sigma^A - \Sigma^R + \Sigma^K}{2},
\end{equation}
which differs from Ref.\ \cite{morimoto_topological_2016} by opposite definitions of $\Sigma^R$ and $\Sigma^A$. Furthermore, for an ideal bath, we have
\begin{align}
    H_B & = \sum_{\bm k,j} \xi_j d_{\bm kj}^\dag d_{\bm kj}, \\
    H_{SB} & = \lambda \sum_{\bm k} \sum_{a,j} (c_{\bm ka}^\dag d_{\bm kj} + d_{\bm kj}^\dag c_{\bm ka}),
\end{align}
where $j$ are bath degrees of freedom, $\xi_j = \varepsilon_j - \mu$ with $\mu$ the chemical potential, and $\lambda$ gives the coupling between the system and the bath. Here $c_{\bm kj}^\dag$ ($c_{\bm kj}$) creates (destroys) a system particle and $d_{\bm kj}^\dag$ ($d_{\bm kj}$) creates (destroys) a bath particle which can be fermionic or bosonic with $[d_{\bm kj}, d_{\bm k'j'}^\dag]_\pm = \delta_{\bm k \bm k'} \delta_{jj'}$ where $+/-$ is the commutator/anticommutator. The total Hamiltonian is
\begin{equation}
    H(t) = H_S(t) + H_B + H_{SB},
\end{equation}
where $H_S(t)$ is given by Eq.\ \eqref{eq:Hsys}. The retarded and advances self energies are given by \cite{jauho_time-dependent_1994}
\begin{equation}
    \left[ \Sigma^{R/A}(t,t') \right]_{ab} = \sum_{j,j'} ( H_{SB} )_{aj} G^{R/A}_{B,jj'}(t,t') ( H_{SB}^\dag )_{j'b},
\end{equation}
where
\begin{align}
    G_B^R(\bm kj; t - t') & = - i \Theta( t - t') \langle [ d_{\bm kj}(t), d_{\bm kj}^\dag(t') ]_\pm \rangle \\
    & = - i \Theta( t - t') e^{-i \xi_j (t - t')}, \\
    G_B^A(\bm kj; t - t') & = i \Theta( t' - t) \langle [ d_{\bm kj}(t), d_{\bm kj}^\dag(t') ]_\pm \rangle \\
    & = i \Theta( t' - t) e^{-i \xi_j (t - t')},
\end{align}
such that
\begin{align}
    & \left[ \Sigma^R(\bm k, t - t') \right]_{ab} \\
    & = - i \delta_{ab} \lambda^2 \Theta( t - t') \sum_j e^{-i\xi_j(t-t')} \\
    & = - i \delta_{ab} \lambda^2 \Theta( t - t') \int_{-\infty}^\infty \frac{d\omega}{2\pi} \, \rho(\omega + \mu) e^{-i\omega(t-t')} \\
    & \approx - \frac{i \Gamma}{2} \, \delta_{ab} \delta(t - t'), 
\end{align}
where we used $\Theta(0) = 1/2$ and we assumed a constant density of states $\rho(\omega) \approx \rho_0$ (the wide-band limit of the bath), with $\Gamma = \lambda^2 \rho_0$. Similarly, one finds $\left[ \Sigma^A(\bm k, t - t') \right]_{ab} \approx (i \Gamma / 2) \delta_{ab} \delta(t - t')$. In frequency space, this yields Eq.\ \eqref{eq:selfenergy-bose} and Eq.\ \eqref{eq:selfenergy-fermi} of the main text.

\section{Analytical results}

\subsection{Frequency integral} \label{app:integral}

The $\omega$ integral in Eq.\ \eqref{eq:jband_final} can be solved by closing the counter in the upper complex plane. 
Indeed, for $\text{Re} \, z > 0$ the integrand decays as $e^{-\beta \, \text{Re} \, z}$ for both the fermionic and bosonic bath, such that the contribution of the upper great half circle vanishes.

\subsubsection{Fermionic bath}

For the fermionic bath, the frequency integral gives
\begin{align}
    & \int_{-\infty}^\infty \frac{d\omega}{2\pi i} \, \frac{f_-^0(\omega)}{\left( \omega - a - i \Gamma/2 \right) \left( \omega - b + i \Gamma/2 \right)} \\
    & = \frac{f_-^0(a + i \Gamma/2)}{a - b + i \Gamma} \\
    & - \frac{1}{\beta} \sum_{j=0}^\infty \frac{1}{\left( z_j - a - i \Gamma/2 \right) \left(z_j - b + i \Gamma/2 \right)},
\end{align}
with $z_j = \mu + i \pi ( 2j + 1 ) / \beta$ and where $a$ and $b$ in Eq.\ \eqref{eq:jband_final} correspond to Floquet eigenenergies when we plug in the single-particle Lehmann representation of the Green's functions. The sum can be evaluated, e.g., with Mathematica, and we find
\begin{equation}
    \frac{1}{a - b + i\Gamma} \left[ f_-^0(a + i \Gamma/2) + \frac{\psi \left( \frac{1}{2} + A \right) - \psi \left( \frac{1}{2} + B \right)}{2\pi i} \right],
\end{equation}
where $\psi(z)$ is the digamma function, $A = i \beta \left( a - \mu + i \Gamma / 2 \right) / 2\pi$, and $B = i \beta \left( b - \mu - i \Gamma / 2 \right) / 2\pi$. Making use of 
\begin{align}
    f_-^0(\omega) & = \frac{1}{2} \left\{ 1 - \tanh \left[ \frac{\beta}{2} \left( \omega - \mu \right) \right] \right\} \\
    & = \frac{1}{2} + \frac{\psi \left( \frac{1}{2} - \gamma \right) - \psi \left( \frac{1}{2} + \gamma \right)}{2\pi i},
\end{align}
with $\gamma = i\beta \left( \omega - \mu \right) / 2 \pi$, the fermionic integral \cite{matsyshyn_fermi-dirac_2023}
\begin{widetext}
\begin{align}
    & \frac{1}{a - b + i\Gamma} \left\{ \frac{1}{2} + \frac{\psi \left[ \frac{1}{2} - \frac{i\beta}{2\pi} \left( a - \mu + \frac{i \Gamma}{2} \right) \right] - \psi \left[ \frac{1}{2} + \frac{i\beta}{2\pi} \left( b - \mu - \frac{i \Gamma}{2} \right) \right]}{2\pi i} \right\} \\
    & = \frac{1}{a - b + i\Gamma} \left[ \frac{\theta(\mu - a) + \theta(\mu - b)}{2} - \ln \left| \frac{a - \mu}{b - \mu} \right| + \frac{\Gamma}{4\pi} \left( \frac{1}{a-\mu} + \frac{1}{b-\mu} \right) + \mathcal O \left( \beta^{-2} \right) \right],
\end{align}
where the last line is a low-temperature expansion with respect to $|a - \mu|$ and $|b-\mu|$ while keeping $\beta \Gamma$ constant. Note that the temperature dependence only enters in the subleading terms.
\end{widetext}

\subsubsection{Bosonic bath}

For the bosonic bath, the frequency integral gives
\begin{align}
    & \mathcal P \int_{-\infty}^\infty \frac{d\omega}{2\pi i} \, \frac{f_+^0(\omega)}{\left( \omega - a - i \Gamma/2 \right) \left( \omega - b + i \Gamma/2 \right)} \\
    & = \frac{f_+^0(a + i \Gamma/2)}{a - b + i \Gamma} + \frac{1}{2 \beta a b} \\
    & + \frac{1}{\beta} \sum_{j=1}^\infty \frac{1}{\left( z_j - a - i \Gamma/2 \right) \left(z_j - b + i \Gamma/2 \right)},
\end{align}
with $z_j = \mu + i 2 \pi j / \beta$ and $\mathcal P$ indicates the Cauchy principal value. Here we took into account that the pole of $f_+^0$ at the origin only contributes half of its residue since it lies on the contour. The sum can again be evaluated with Mathematica and we find
\begin{equation}
    \begin{aligned}
        & \frac{1}{a - b + i\Gamma} \left[ f_+^0(a + i \Gamma/2) - \frac{\psi \left( 1 + A \right) - \psi \left( 1 + B \right)}{2\pi i} \right] \\
        & + \frac{1}{2\beta ab}.
    \end{aligned}
\end{equation}
Making use of 
\begin{align}
    f_+^0(\omega) & = \frac{1}{2} \left\{ \coth \left[ \frac{\beta}{2} \left( \omega - \mu \right) \right] - 1 \right\} \\
    & = \frac{1}{z} - \frac{1}{2} + \frac{\psi \left( 1 + \gamma \right) - \psi \left( 1 - \gamma \right)}{2\pi i},
\end{align}
the bosonic integral becomes
\begin{widetext}
\begin{equation}
    \frac{-1}{a - b + i\Gamma} \left\{ \frac{1}{2} + \frac{\psi \left[ 1 - \frac{i\beta}{2\pi} \left( a - \mu + \frac{i \Gamma}{2} \right) \right] - \psi \left[ 1 + \frac{i\beta}{2\pi} \left( b - \mu - \frac{i \Gamma}{2} \right) \right]}{2\pi i} - \frac{1}{2\beta \left( a - \mu + \frac{i \Gamma}{2} \right)} - \frac{1}{2\beta \left( b - \mu - \frac{i \Gamma}{2} \right)} \right\},
\end{equation}
which up to order $\beta^{-2}$ has the same low-temperature limit, up to an overall minus sign, as the fermionic case. The minus sign cancels with the prefactor in Eq.\ \eqref{eq:jband_final} such that the particle statistics of the ideal bath becomes unimportant at sufficiently low temperature and system-bath coupling $\Gamma$ while keeping $\beta \Gamma$ constant.
\end{widetext}

\subsection{Simple chain} \label{app:1dchain}

For the simple chain in $D=1$, the Floquet Hamiltonian (for a given momentum $k$) is given by
\begin{equation}
    H_F = \sum_l l \Omega \left| \phi_{kl} \right> \left< \phi_{kl} \right|,
\end{equation}
where the projector $\left( \left| \phi_{kl} \right> \left< \phi_{kl} \right| \right)_{mn} = e^{-ik(m-n)a} J_{l-m}(\zeta) J_{l-n}(\zeta)$ with $\zeta=2J/\Omega$. We further have
\begin{equation}
    G^{R/A}(k,\omega) = \sum_l \frac{\left| \phi_{kl} \right> \left< \phi_{kl} \right|}{\omega - l \Omega \pm i \Gamma / 2},
\end{equation}
and 
\begin{equation}
    \partial_k H_F = \sum_l l\Omega \left( \left| \partial_k \phi_{kl} \right> \left< \phi_{kl} \right| + \left| \phi_{kl} \right> \left< \partial_k \phi_{kl} \right| \right).
\end{equation}
Hence, the current for the ideal fermionic ($+$) or bosonic ($-$) bath becomes
\begin{widetext}
\begin{align}
    j & = \pm e \Gamma \sum_{m,n} \sum_{l,l'} \int_{\omega,k} f_\pm^0(\omega - n \Omega) \, \frac{\left( \partial_k H_F \right)_{0m} \left( \left| \phi_{kl} \right> \left< \phi_{kl} \right| \right)_{mn} \left( \left| \phi_{kl'} \right> \left< \phi_{kl'} \right| \right)_{n0}}{\left( \omega - l \Omega + i \Gamma / 2 \right) \left( \omega - l' \Omega - i \Gamma / 2 \right)} \\
    & = \mp iae \Gamma \sum_{m,n} \sum_{l,l'} \sum_p p \Omega \int_{\omega,k} f_\pm^0(\omega + n \Omega) \, \frac{m J_{p}(\zeta) J_{p+m}(\zeta) J_{l+m}(\zeta) J_{l+n}(\zeta) J_{l'+n}(\zeta) J_{l'}(\zeta)}{\left( \omega - l \Omega + i \Gamma / 2 \right) \left( \omega - l' \Omega - i \Gamma / 2 \right)}. \label{eq:j_chain3}
\end{align}
\end{widetext}
We see explicitly that the ac response vanishes, since this would result in an extra phase factor $e^{ikna}$ in Eq.\ \eqref{eq:j_chain3} for the $n$th harmonic, yielding $\delta_{n0}$ after performing the momentum integral. Using
\begin{equation}
    \sum_n n J_n(\zeta) J_{n+m}(\zeta) = \frac{\zeta}{2} \delta_{m,\pm 1},
\end{equation}
we obtain
\begin{widetext}
\begin{align}
    j & = \mp ieJ \Gamma \sum_n \sum_{l,l'} \int_\omega f_\pm^0(\omega + n \Omega) \, \frac{
    \left[ J_{l+1}(\zeta) - J_{l-1}(\zeta) \right] J_{n+l}(\zeta) J_{n+l'}(\zeta) J_{l'}(\zeta)}{\left( \omega - l \Omega + i \Gamma / 2 \right) \left( \omega - l' \Omega - i \Gamma / 2 \right)} \\
    & = \mp ieJ \Gamma \sum_n \sum_{l,l'} \int_\omega f_\pm^0(\omega) \, \frac{
    \left[ J_{l+1-n}(\zeta) - J_{l-1-n}(\zeta) \right] J_l(\zeta) J_{l'}(\zeta) J_{l'-n}(\zeta)}{\left( \omega - l \Omega + i \Gamma / 2 \right) \left( \omega - l' \Omega - i \Gamma / 2 \right)} \\
    & = \mp ieJ \Gamma \sum_l \int_\omega \frac{f_\pm^0(\omega)J_l(\zeta)}{\omega - l \Omega + i \Gamma / 2} \left( \frac{
    J_{l+1}(\zeta)}{\omega - (l+1) \Omega - i \Gamma / 2} - \frac{
    J_{l-1}(\zeta)}{\omega - (l-1) \Omega - i \Gamma / 2} \right) \\
    & = \mp eJ \Gamma \sum_l J_l(\zeta) J_{l+1} (\zeta) \int_{-\infty}^\infty \frac{d\omega}{2\pi i} \left[ \frac{f_\pm^0(\omega)}{\left( \omega - a - i \Gamma / 2 \right) \left( \omega - b + i \Gamma / 2 \right)} - \frac{f_\pm^0(\omega)}{\left( \omega - b - i \Gamma / 2 \right) \left( \omega - a + i \Gamma / 2 \right)} \right],
\end{align}
\end{widetext}
where we used $\sum_n J_n(\zeta) J_{n+m}(\zeta) = \delta_{m0}$ in the second line, $a = l \Omega$, and $b = (l+1) \Omega$. Using our result above for the frequency integral, we obtain for the fermionic bath
\begin{widetext}
\begin{align}
    j & = -2eJ \Gamma \sum_l J_l(\zeta) J_{l+1} (\zeta) \, \text{Re} \left( \frac{1}{\Omega - i \Gamma} \frac{\psi \left[ \frac{1}{2} - \frac{i\beta}{2\pi} \left( l \Omega - \mu + \frac{i\Gamma}{2} \right) \right] - \psi \left[ \frac{1}{2} + \frac{i\beta}{2\pi} \left( (l+1) \Omega - \mu - \frac{i\Gamma}{2} \right) \right]}{2\pi i} \right) \label{eq:j_chain_exact} \\
    & = \frac{1}{\pi} \frac{eJ\Omega/\Gamma}{1 + ( \Omega / \Gamma )^2} \sum_l J_l(\zeta) \left\{ \left[ J_{l-1}(\zeta) + J_{l+1}(\zeta) \right] \text{Im}\left( Q_l \right) + \frac{\Gamma}{\Omega} \left[ J_{l-1}(\zeta) - J_{l+1}(\zeta) \right] \text{Re}\left( Q_l \right) \right\} \\
    & = \frac{1}{\pi} \frac{2eJ\Omega/\Gamma}{1 + ( \Omega / \Gamma )^2} \sum_l J_l(\zeta) \left[ \frac{l J_l(\zeta)}{\zeta} \, \text{Im}\left( Q_l \right) + \frac{\Gamma}{\Omega} \frac{dJ_l(\zeta)}{d\zeta} \, \text{Re}\left( Q_l \right) \right], \label{eq:j_chain_exact2}
\end{align}
\end{widetext}
with $Q_l \equiv \psi \left[ 1/2 + (i \beta / 2 \pi ) \left( l \Omega - \mu + i\Gamma / 2 \right) \right]$ and $\zeta = 2J/\Omega$. A similar result can be obtained for the bosonic bath. We find numerically that the part of the sum that is proportional to $\text{Im} \left( Q_l \right)$ gives the largest contribution and depends weakly on $\Omega$, while the other part that is proportional to $\text{Re} \left( Q_l \right)$ is relatively smaller and gives rise to oscillations. Moreover, note that 1D inversion symmetry ($x \mapsto -x$) implies that $j(-\mathcal E) = -j(\mathcal E)$ and which can be seen explicitly in Eq.\ \eqref{eq:j_chain_exact2} as $J_{-l}(-\zeta) = J_l(\zeta)$.
\begin{figure*}
    \centering
    \includegraphics[width=.8\linewidth]{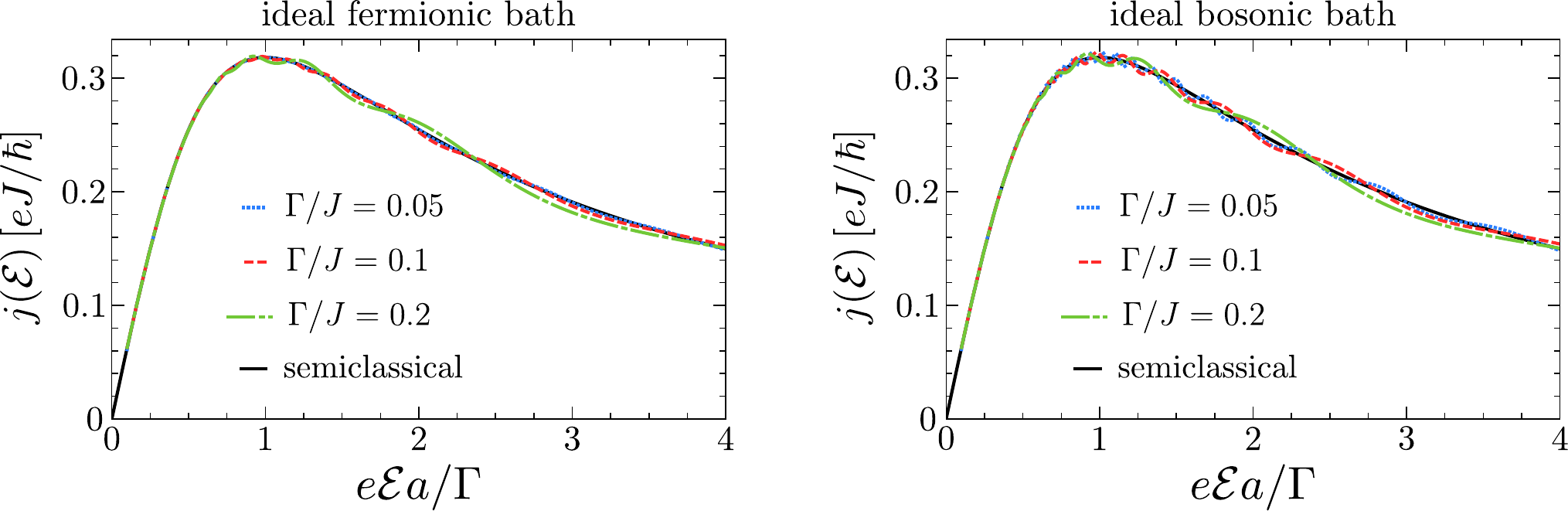}
    \caption{\textbf{Exact results for the simple chain coupled to an ideal bath.} Steady-state current for the simple chain as given in Eq.\ \eqref{eq:j_chain_exact} calculated for $l \in [-400,400]$ at half filling with $J \beta = 50$, and different values of $\Gamma$ as indicated. The solid line gives the semiclassical result [Eq.\ \eqref{eq:j_chain_sc}].}
    \label{fig:j_chain_exact}
\end{figure*}

\bibliography{references}

\end{document}